\documentclass[twocolumn]{pasj00}

\begin{document}

\SetRunningHead{Several AGNs}{H. Noda et al.}
\title{The Nature of  the Stable Soft X-ray Emissions \\in Several 
Types of Active Galactic Nuclei Observed by Suzaku}

\author{Hirofumi \textsc{Noda},\altaffilmark{1}
Kazuo \textsc{Makishima},\altaffilmark{1,2,3}
Kazuhiro \textsc{Nakazawa},\altaffilmark{1} 
Hideki \textsc{Uchiyama},\altaffilmark{1} \\
Shin'ya \textsc{Yamada},\altaffilmark{2}
and Soki \textsc{Sakurai}\altaffilmark{1}
}
\altaffiltext{1}{
  Department of Physics, University of Tokyo\\
  7-3-1, Hongo, Bunkyo-ku, Tokyo, 113-0033, Japan}
\altaffiltext{2}{Institute of Physical and Chemical Research (RIKEN)\\
  2-1 Hirosawa, Wako-shi, Saitama 351-0198}
  \altaffiltext{3}{Research Center for the Early Universe, University of Tokyo\\
   7-3-1, Hongo, Bunkyo-ku, Tokyo, 113-0033, Japan}

\email{noda@juno.phys.s.u-tokyo.ac.jp}

\Received{2012 May 29}
\Accepted{2012 August 15}
\Published{$\langle$publication date$\rangle$}

\maketitle

\begin{abstract}
To constrain the origin of the soft X-ray excess phenomenon
seen in many active galactic nuclei, 
the intensity-correlated spectral analysis,
developed by Noda et al. (2011b) for Markarian 509,
was applied to wide-band (0.5--45 keV)
Suzaku data of five representative objects with relatively weak reflection signature.
They are the typical bare-nucleus type 1 Seyfert  Fairall 9,
the bright and typical type 1.5 Seyfert MCG-2-58-22,  
3C382 which is one of the X-ray brightest broad line radio galaxies,
the typical Seyfert-like radio loud quasar 4C+74.26, 
and the X-ray brightest radio quiet quasar MR2251-178. 
In all of them, soft X-ray intensities in energies below 3 keV 
were tightly correlated with that in 3--10 keV, 
but with significant positive offsets. 
These offsets, when calculated in finer energy bands, 
define a stable soft component in 0.5--3 keV. 
In each object, this component successfully explained the soft excess above a power-law fit.  
These components were interpreted in several alternative ways, 
including a thermal Comptonization component which is 
independent of the dominant power-law emission. 
This interpretation, considered physically most reasonable, is discussed from a viewpoint 
of Multi-Zone Comptonization, 
which was proposed for the black hole binary Cygnus X-1 (Makishima et al. 2008). 
\end{abstract}

\section{Introduction}
\label{sec:intro}

In soft X-ray spectra of Active Galactic Nuclei (AGNs), 
a phenomenon called ``soft X-ray excess'' is often observed.
Characterized by a similar and steep flux upturn towards lower energies, 
this feature is clearly noticed particularly in weakly-absorbed AGNs, 
such as type I Seyferts and Broad Line Radio Galaxies (BLRG).
For years, the origin of this spectral structure has been unidentified,
and many interpretations have been proposed. 
Its simplest explanation is 
blackbody radiation from an optically-thick accretion disk.
However, the observed color temperature of the soft excess, typically $\sim 0.2$ keV, 
is too high for such disks around black holes (BHs) 
of  $\gtrsim 10^7~M_\odot$ in mass, where $M_\odot$ is the solar mass.


\begin{table*}[t]
 \caption{Information of the five AGNs to be studied.}
 \label{all_tbl}
 \begin{center}
  \begin{tabular}{ccccccc}
   \hline\hline
  Object name  &Type & Obs. date & redshift &$N_{\rm H}$(Gal)$^{*}$  & reported $R^{\dagger}$ & previous Suzaku study  \\

   \hline
  Fairall 9    	    & Sy1 	&   2007 June 7	& $0.047$& 0.031   	& 0.52$^{+0.20}_{-0.18}$	& Patrick et al. (2011)\\
  			   &	&   2010 May 19	&		&	&1.55$^{+0.26}_{-0.24}$  & Patrick et al. (2011) \\
 MCG-2-58-22     & Sy1.5	&    2009 December 2&   $0.047$ &	0.027 	& 	$0.69\pm0.05$	& Rivers et al. (2011)\\
  3C382    		  & BLRG  &    2007 April 27	&$0.058$	&  0.074	& 0.15$\pm$0.05 	& Sambruna et al. (2011)\\
  4C+74.26        & RLQ	  &    2007 October 25	&$0.104$& 	0.119 	& 	0.3--0.7	& Larsson et al. (2008)\\
  MR2251-178 	   & RQQ &   2009 May 7& $0.064$	& 	0.024 	& 	$\leq$0.2	 & Gofford et al. (2011)\\
                      
      \hline\hline

  \end{tabular}
 \end{center} 
      
            	{\small
	\footnotemark[$*$] Equivalent hydrogen column density of the Galactic line-of-sight absorption in  $10^{22}$ cm$^{-2}$.  \\
         \footnotemark[$\dagger$] Reflection fraction defined by  $R = \Omega /2 \pi$, when $\Omega$ is the reflection solid angle.}

\end{table*}

Given this, 
various alternative interpretations were proposed. 
Among them, mainly three ideas have been promising.
One is absorption by  a partially-covering and ionized absorber 
often incorporating velocity smearing effects 
(e.g., Middleton \& Done 2004; Schurch \& Done 2008; O'Neil et al. 2007), 
another is relativistically blurred reflection from an ionized disk which is in similar conditions as the first case 
(Zoghbi et al. 2010; Nardini et al. 2011), 
and the other is a thermal Comptonization component that is separated from
that producing the dominant Power-Law (PL) component (Marshall et al. 2003).  
Since these interpretations often degenerate in spectral analysis (Cerruti et al. 2011), 
there have been no conclusions for the origin of the soft X-ray excess in AGNs. 
Recently, however, Makishima et al. (2008) and Yamada (2011)  
established a Multi-Zone Comptonization (MZC) view for Cygnus X-1 (hereafter Cyg X-1) 
through the analysis of 0.5--200 keV Suzaku data, 
and showed that the last interpretation among the three indeed explains 
the soft excess (Frontera et al. 2001) seen in this leading black hole binary (BHB). 
This is analogous to the third interpretation of 
the AGN soft excess as described above. 

To examine the origin of the soft X-ray excess in AGNs utilizing the wide-band Suzaku capability, 
Noda et al. (2011b) chose the typical weakly-absorbed type 1 Seyfert galaxy Markarian 509 (Mrk 509 hereafter), 
and developed a method to study how its 0.5--3 keV intensity 
in the Suzaku data is correlated to 
that in 3--10 keV, and found that a significant positive offset remains in the softer band intensity 
when the correlation is extrapolated to lower counts.  
Because the method is based on Count-Count Correlation with Positive Offset, 
hereafter, it is called ``C3PO'' method for simplicity.  
Thus, utilizing time variations with the C3PO method, 
they extracted a stable soft X-ray component, and successfully explained it by 
a thermal  Comptonization which is independent of the dominant PL continuum. 
When this soft Comptonization component is included, 
the time-averaged 0.5--45 keV spectrum of Mrk 509 was reproduced 
in terms of a weakly absorbed single PL and its reflection.  
In addition, Noda et al. (2011b) discovered that the new soft component varied on a  timescale longer  
than 3 days, independently of the PL variation.
This securely excluded a competing interpretation of the detected soft component in terms of 
some largely extended thermal emission from the host galaxy.

At the same time as Noda et al. (2011b), 
Mehdipour et al. (2011) analyzed multi-wavelength data of the same object, Mrk 509, 
which were  obtained in a large campaign including XMM-Newton, Hubble, and FUSE.
They obtained the same conclusion, that the soft X-ray excess of this AGN is 
created as a thermal Comptonization component which is independent of the principal PL continuum. 
Importantly, the parameters of the thermal Comptonization they derived, including
an electron temperature of $\sim 0.2$ keV, 
and an optical depth of $\sim 16$, agree with those of Noda et al. (2011b).
Thus, through the two independent methods 
(timing analysis and  multi-wavelength spectral fitting),  
the soft excess phenomenon in Mrk 509 has been confirmed to arise as Comptonization 
by a warm corona, which is presumably different from (though possibly related to) the hotter corona 
producing the harder PL continuum. 
When combined with the Suzaku results on Cyg X-1 quoted above, 
this result strengthens the general analogy between 
BHBs and AGNs. 
 
The next step is to examine whether or not the same phenomenon as found in Mrk 509 and Cyg X-1
is present in a larger number of AGNs, including Seyferts and objects of other types.  
For this purpose, we utilize Suzaku archival datasets of AGNs, 
because the wide-band simultaneous coverage with Suzaku, typically available over a 0.5--45 keV band, 
is essential in determining the underlying continuum and the reflection component for each AGN, 
and hence to unambiguously identify the soft excess signals. 
In fact, many AGNs have been observed with Suzaku since its launch in 2005. However,  
the overall sample is neither complete nor homogeneous. 
Therefore, we choose to conduct the present study via the following 
two steps. The first is to define those AGN types which are suitable to 
our purpose.  The second is to select, from the Suzaku archive, the best
target that represents each of the selected classes.

In the present work, 
errors refer to $\pm 1\sigma$ confidence limits, except model-fitting parameters in XSPEC
for which $90\%$ error ranges are adopted. 

\section{Target Selection}
\label{sec:selection}
\begin{figure*}[t]
 \begin{center}
  \FigureFile (150mm,150mm)
    {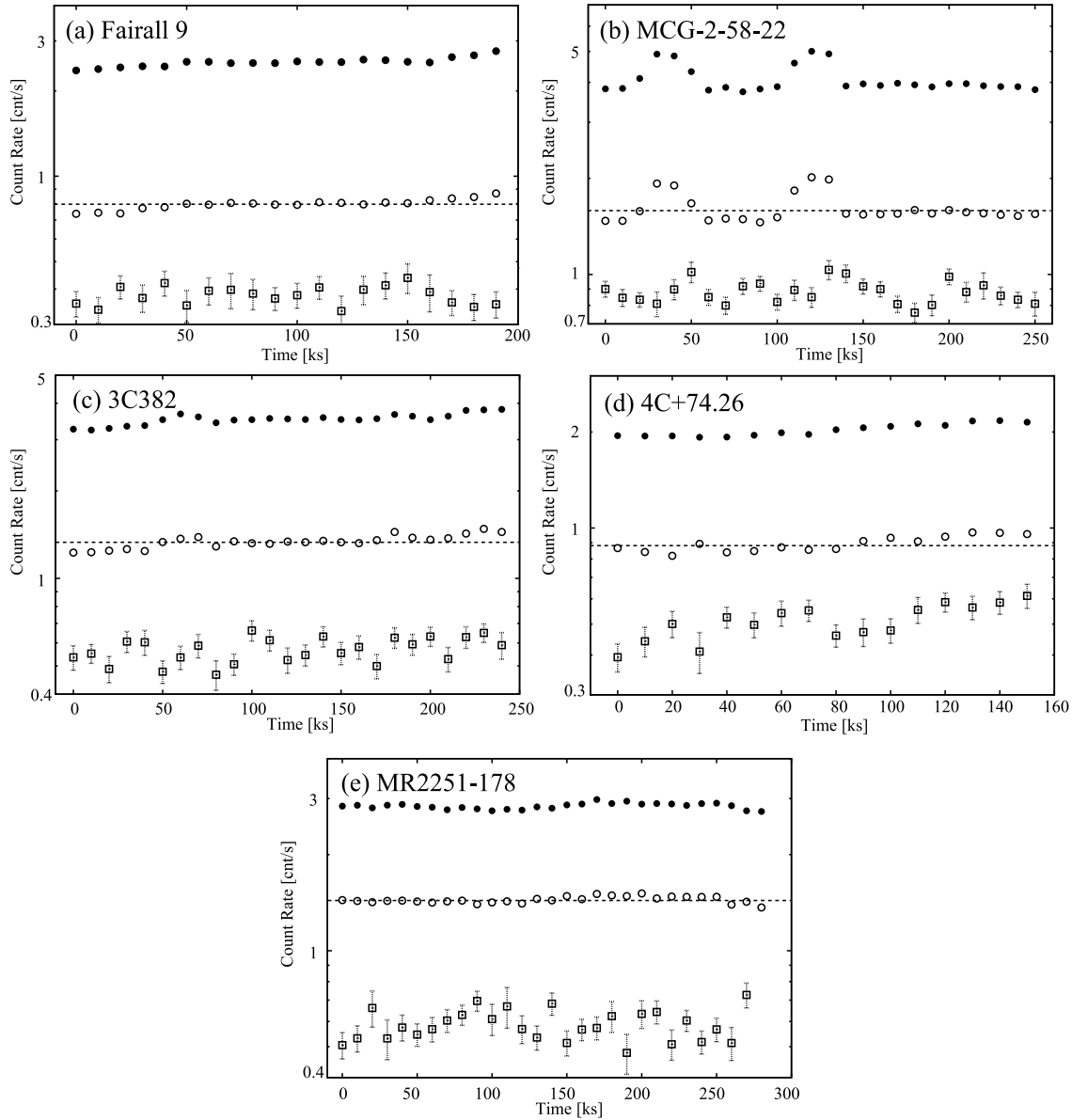}
 \end{center}
\setlength{\belowcaptionskip}{0mm}
 \caption
{Background-subtracted and dead-time corrected
light curves of the five AGNs 
measured with  XIS FI, in 0.5--3.0 keV represented by filled circles and 3--10 keV by open circles, 
shown with a binning of 10 ks. 
Open squares represent those of HXD-PIN in 15--45 keV, multiplied by a factor of 5 and with the same binning. 
$1\sigma$ errors of the data points of XIS FI are all less than 0.04 cnt s$^{-1}$. 
Dotted lines show the average 3--10 keV count rates.}
\label{fig:lc}
\end{figure*}

In order for the C3PO method developed in Noda et al. (2011b) to be applicable, 
the target AGNs should satisfy several conditions.  
First, we should exclude jet dominated systems like Blazars.  
This condition enables us to treat pure X-ray emission from a close vicinity of the central black hole (BH).  
Second, absorptions in our Galaxy and the host galaxies of AGNs should both be low enough to 
detect soft X-ray signals with significant statistics. 
Finally, reflection from accretion disks and other surrounding matters should be weak: 
our C3PO method utilizes count rates in the 3--10 keV 
bands as references, so that they should be dominated by the continuum signals rather than reflection. 
Taking these conditions into account, 
we have chosen five AGN types; type 1 Seyfert galaxies (Sy1), type 1.5 Seyfert galaxies (Sy1.5),  
Broad Line Radio Galaxies (BLRGs), Seyfert-like Radio Loud Quasars (RLQs), and Radio Quiet Quasars (RQQs). 

\begin{figure*}[t]
 \begin{center}
   \FigureFile(145mm,145mm)
    {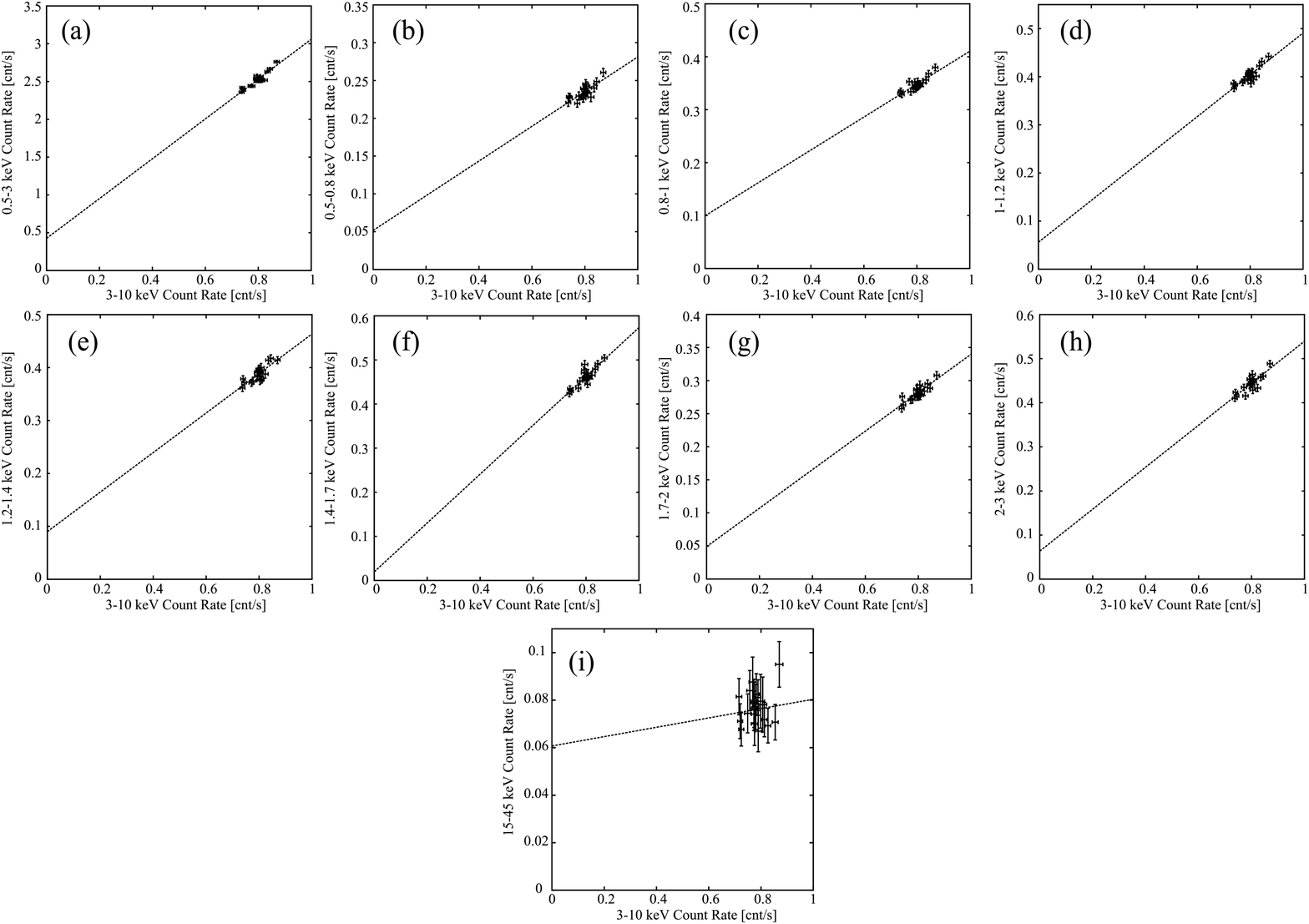}
 \end{center}
\setlength{\belowcaptionskip}{0mm}
 \caption
{The CCPs of Fairall 9. Abscissas gives NXB-subtracted XIS FI count rate in 3--10 keV, 
while ordinate gives that in (a) 0.5--3 keV, (b) 0.5--0.8 keV,
(c) 0.8--1.0 keV, (d) 1.0--1.2 keV,
(e) 1.2--1.4 keV, (f) 1.4--1.7 keV, (g)1.7--2.0 keV,
(h) 2.0--3.0 keV, and (i) 15--45 keV. All data are binned into 10 ks. 
The error bars represent statistical $\pm1\sigma$ range. The dotted straight line refers to equation (1).}
\label{fig:ccp}
\end{figure*}

As representatives of Sy 1, Sy 1.5, BLRGs, RLQs, and RQQs, we chose Fairall 9, MCG-2-58-22, 
3C382, 4C+74.26, and MR2251-178, respectively. 
Fairall 9, 3C382, and MR2251-178 are ones of the most typical and brightest objects 
among the currently available Suzaku datasets of respective types of AGNs. 
MCG-2-58-22 was selected because of its brightness, moderate absorption,   
and the longest net exposure, among the available Suzaku datasets of Sy1.5 (Rivers et al. 2011). 
4C+74.26 was chosen because  its X-ray spectrum is similar to those of  Seyferts, and 
jet contribution to its 0.5--45 keV signal is low (Kataoka et al. 2011). 
These AGNs were commonly reported to have mild reflection components, 
e.g., by Schmoll et al. (2009), Rivers et al. (2011), Sambruna et al. (2011), Larsson et al. (2008) and Gofford et al. (2011),  
who analyzed the same Suzaku datasets as we utilize.  
We summarize these objects in table 1. 
The Suzaku archive provides two datasets of Fairall 9, obtained in 2007 and 2010. 
Among them,  we chose the first one for our analysis, 
because the object exhibited on this occasion a  lower reflection fraction and a higher 3--10 keV flux  
than in the other.

\section{Data Reduction}
\label{sec:obs}

From the Suzaku (Mitsuda et al. 2007) archive, we retrieved the XIS (Koyama et al. 2007) and 
HXD (Takahashi et al. 2007) data of the 5 objects selected in section 2. 
These data, all acquired at the XIS nominal position, were prepared via version 2.0 processing in 2005, 
version 2.1 in 2007, and version 2.4 in 2009. 

In the present analysis, 
the data of XIS 0, 2 and 3 (after 2006, XIS 0 and 3 only), which use front-illuminated (FI) CCD chips, 
were added and used as XIS FI. 
The data from XIS 1 were not analyzed, since it uses a back-illuminated CCD, and has a relatively high 
and unstable background. 
On-source events of the three or two XIS FI cameras were extracted from a circular region of $120''$
radius centered on the source.
Background events were accumulated on a surrounding annular region 
of the same camera, with the inner and outer radii of $180''$ and $270''$, respectively.
The response matrices and ancillary response files were produced  
by \texttt{xisrmfgen} and \texttt{xissimarfgen} (Ishisaki et al. 2007), respectively.

In a similar way, we prepared events of HXD-PIN.
Non X-ray Background (NXB) contained  in the data was estimated by analyzing a set of  fake events
which were created by a standard NXB model (Fukazawa et al. 2009).
The on-source events and the NXB events were analyzed in the same manner,
and the latter was subtracted from the former.
In addition, the contribution from Cosmic X-ray Background (CXB; Boldt et al. 1987) 
was estimated and also subtracted from the on-source data. This was conducted  
using the HXD-PIN response to diffuse sources, 
assuming the spectral CXB surface brightness model
determined by HEAO 1 (Gruber et al. 1999):
$9.0 \times 10 ^{-9} (E/3~\mathrm{keV})^{-0.29} \exp (-E/40~\mathrm{keV})$
erg cm$^{-2}$ s$^{-1}$ str $^{-1}$ keV$^{-1}$,
where $E$ is the photon energy.
The estimated CXB count rate amounts to  5\% of the NXB signals, 
in agreement with Kokubun et al. (2007).

\section{Data Analysis and Results}

\subsection{Extraction of soft excess component}

\begin{figure*}[p]
 \begin{center}
   \FigureFile(145mm,145mm)
    {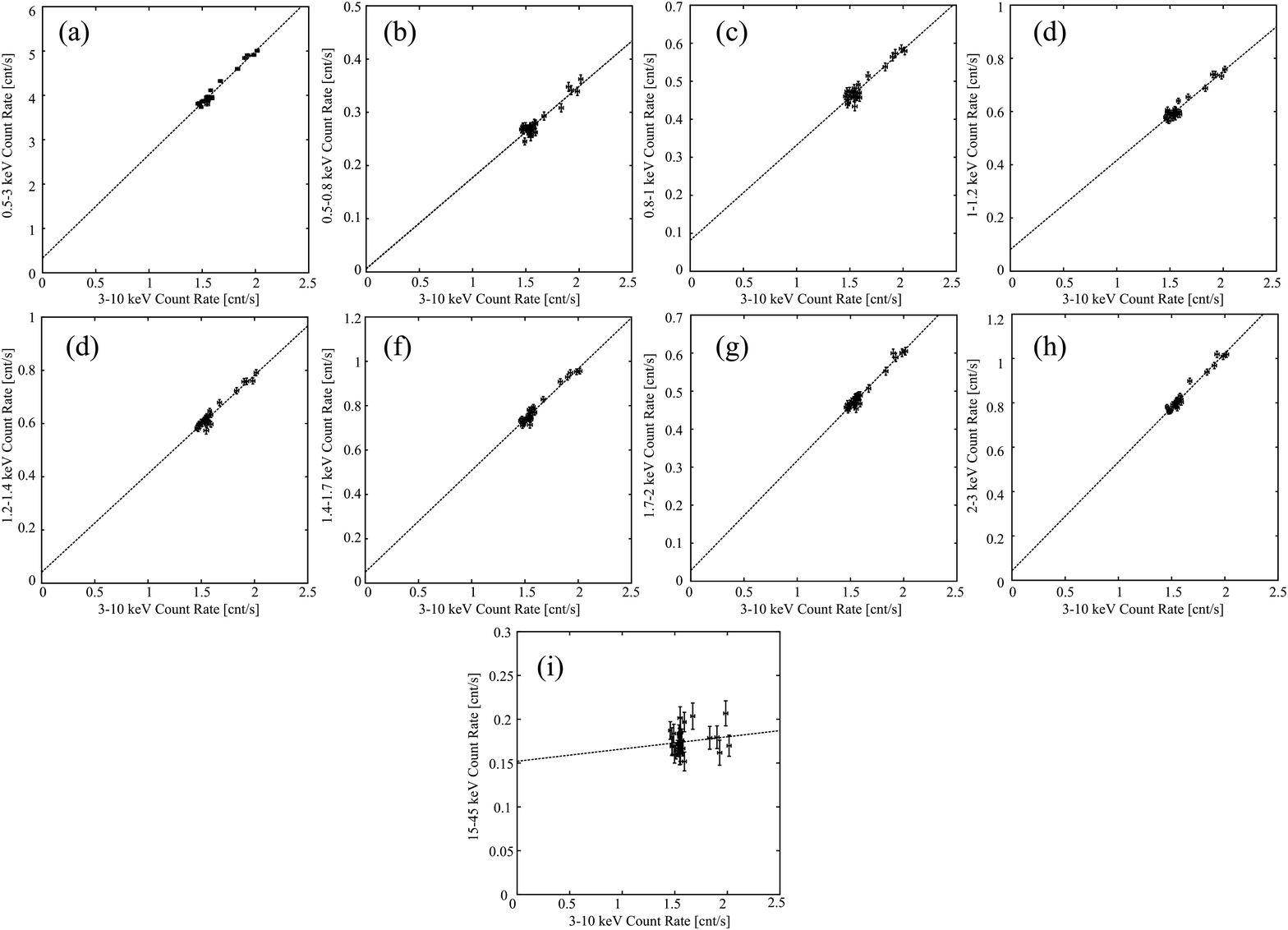}
 \end{center}
\setlength{\belowcaptionskip}{0mm}
 \caption
{The same as figure 3, but of MCG-2-58-22.}
\label{fig:ccp}
\end{figure*}

\begin{figure*}[bthp]
 \begin{center}
   \FigureFile(145mm,145mm)
    {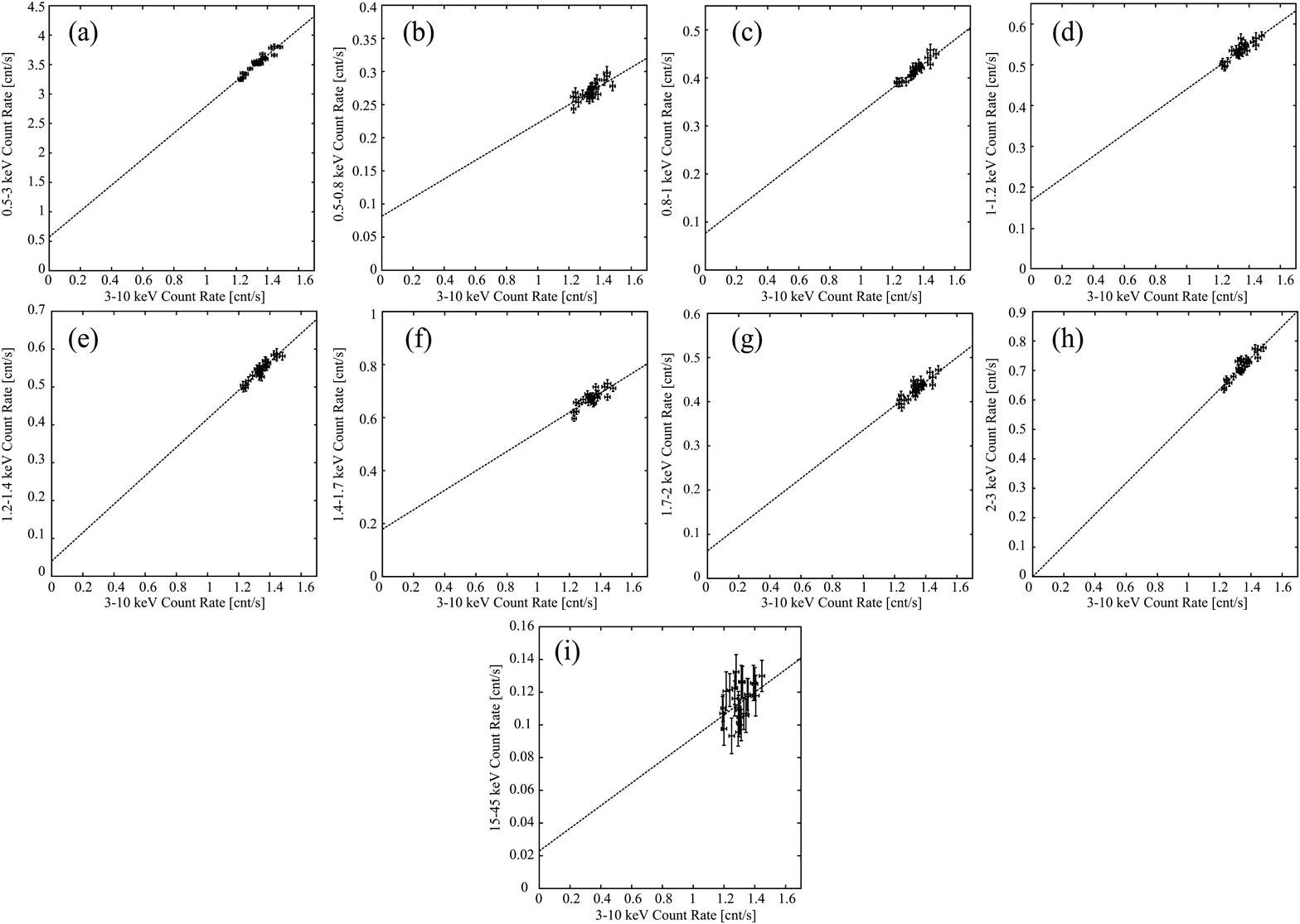}
 \end{center}
\setlength{\belowcaptionskip}{0mm}
 \caption
{The same as figure 3, but of 3C382.}
\label{fig:ccp}
\end{figure*}

\begin{figure*}[ptbh]
 \begin{center}
   \FigureFile(145mm,145mm)
    {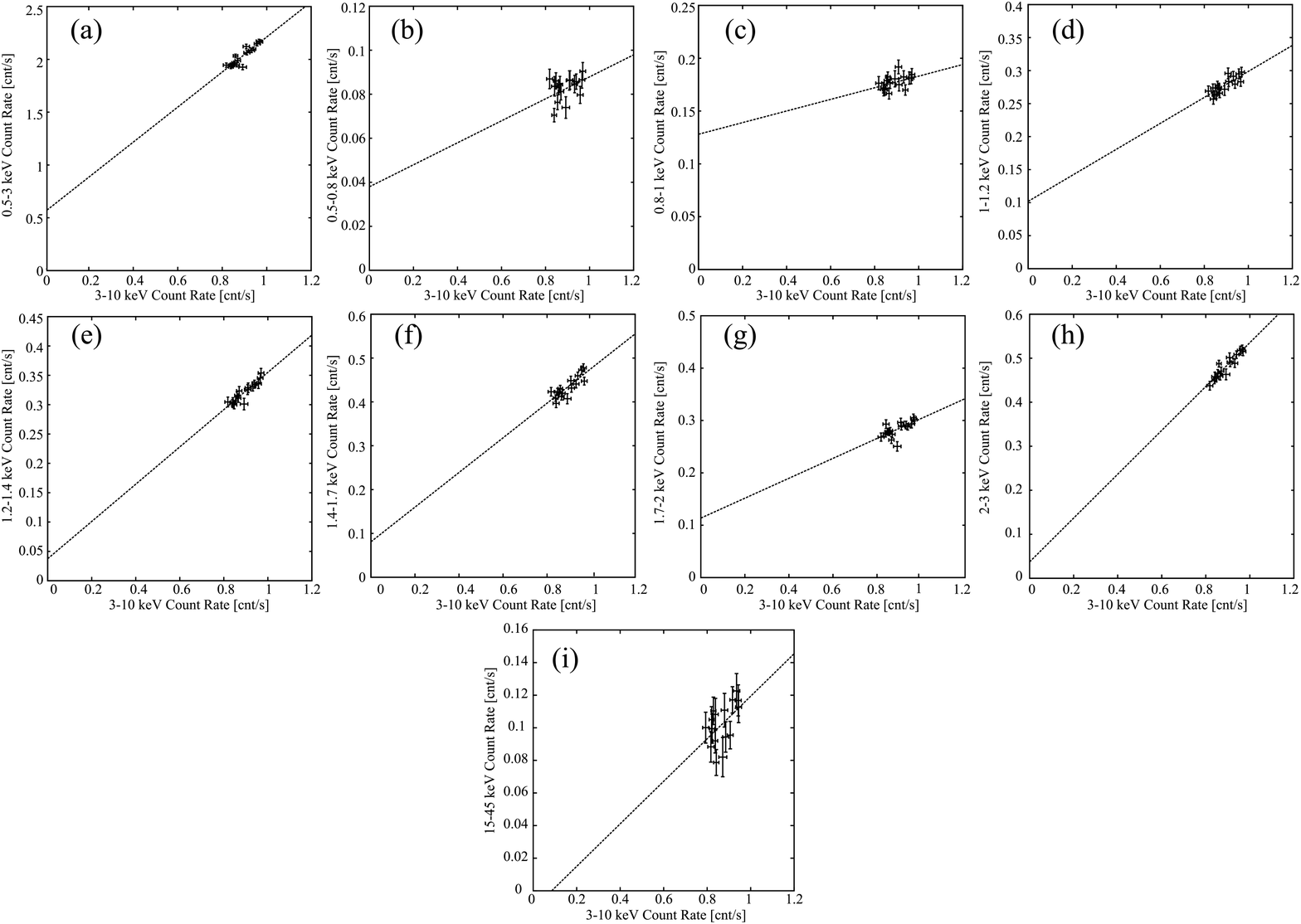}
 \end{center}
\setlength{\belowcaptionskip}{0mm}
 \caption
{The same as figure 3, but of 4C+74.26.}
\label{fig:ccp}
\end{figure*}

\begin{figure*}[bthp]
 \begin{center}
   \FigureFile(145mm,145mm)
    {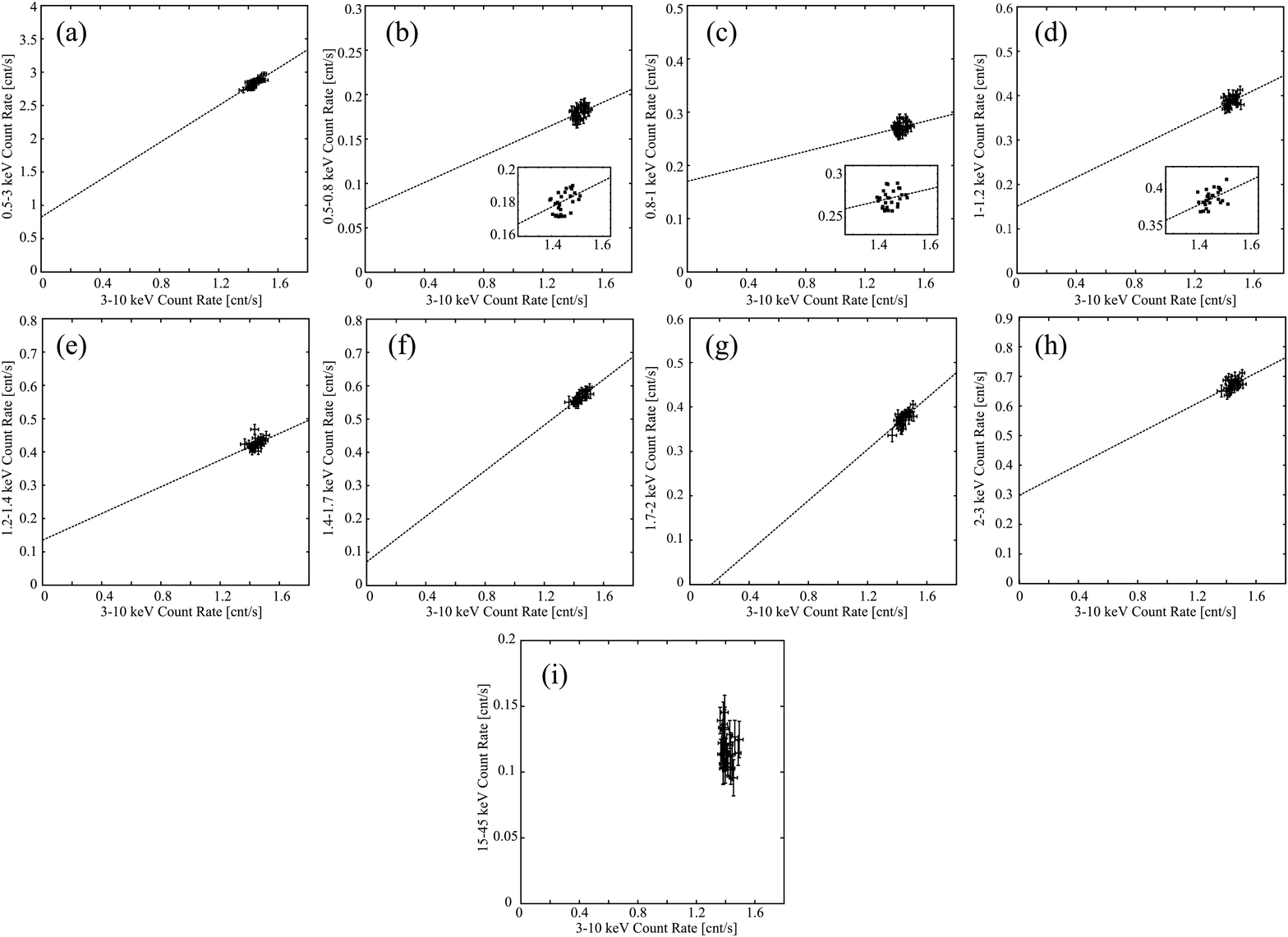}
 \end{center}
\setlength{\belowcaptionskip}{0mm}
 \caption
{The same as figure 3, but of MR2251-178.
Plots without error bars are also shown as insets in panels (b--d). 
The correlation in panel (i) is unconstrained. }
\label{fig:ccp}
\end{figure*}

\renewcommand{\arraystretch}{1}
\begin{table*}[t]
 \caption{Parameters obtained by fitting 7 CCPs in figure 2(a--i) and figure 3-6(a--h) of the five individual AGNs with equation (1).$^{*}$}
 \label{all_tbl}
 \small
 \begin{center}
  \begin{tabular}{ccccc}
   \hline\hline
  Object&Energy range (keV)& $$a & $b$~(counts s$^{-1}$)& $\chi^2 /$d.o.f. \\

   \hline
  Fairall 9 & 0.5--3 & $4.36\pm0.10$ & $0.51\pm0.21$ & 19.2/18 \\\cline{2-5}
  	&  0.5--0.8   & $0.23\pm0.04$ & $0.05\pm0.03$ & 21.1/18\\
   	 &  0.8--1 & $0.31\pm0.04$ & $0.10\pm0.03$ & 13.5/18\\ 
	  &  1--1.2   & $0.44\pm0.05$ & $0.06\pm0.21$ & 16.0/18\\
	  &  1.2--1.4   & $0.37\pm0.06$ & $0.09\pm0.05$ & 22.4/18\\
	  &  1.4--1.7   & $0.55\pm0.07$ & $0.02\pm0.06$ & 23.9/18\\
	   &  1.7--2    & $0.29\pm0.04$ & $0.05\pm0.03$ & 13.8/18\\
	  &  2--3    & $0.48\pm0.07$ & $0.06\pm0.05$ & 24.6/18 \\ \cline{2-5} 
	  & 15--45 & $0.02\pm0.04$& $0.06\pm0.03$ & 12.3/18\\\cline{2-5}
	  \hline
	  	  
  MCG-2-58-22  & 0.5--3 & $2.33\pm0.08$ & $0.33\pm0.13$ & 31.5/24 \\ \cline{2-5}
  			&  0.5--0.8    & $0.17\pm0.01$ & $0.01\pm0.02$ & 40.8/24\\
  			 &  0.8--1    & $0.25\pm0.02$ & $0.08\pm0.03$ & 39.3/24\\
                      	 &  1--1.2    & $0.34\pm0.02$ & $0.08\pm0.03$ & 44.2/24\\
                        &  1.2--1.4    & $0.37\pm0.02$ & $0.04\pm0.03$ & 33.2/24\\
                       &  1.4--1.7    & $0.46\pm0.02$ & $0.05\pm0.03$ & 41.2/24\\
                        &  1.7--2    & $0.29\pm0.01$ & $0.03\pm0.02$ & 20.8/24\\
                      &  2--3    & $0.49\pm0.02$ & $0.04\pm0.03$ & 29.3/24\\\cline{2-5}
                 	  & 15--45 & $0.01\pm0.02$& $0.15\pm0.03$ & 36.3/24\\\cline{2-5}
                      \hline 
       
   3C382  & 0.5--3 & $3.78\pm0.26$ & $0.65\pm0.23$ & 21.6/23 \\ \cline{2-5}
   			&  0.5--0.8    & $0.14\pm0.02$ & $0.08\pm0.03$ & 25.4/23\\
        			&  0.8--1    & $0.25\pm0.02$ & $0.08\pm0.03$ & 9.3/23\\
                        &  1--1.2   & $0.27\pm0.03$ & $0.17\pm0.04$ & 18.7/23\\
                       &  1.2--1.4    & $0.38\pm0.03$ & $0.04\pm0.04$ & 14.1/23\\
                       &  1.4--1.7    & $0.37\pm0.06$ & $0.18\pm0.08$ & 48.8/23\\
                       &  1.7--2    & $0.27\pm0.03$ & $0.06\pm0.04$ & 22.7/23\\
                       &  2--3   & $0.53\pm0.05$ & $0\pm0.06$ & 23.7/23\\\cline{2-5}
                 	  & 15--45 & $0.07\pm0.03$& $0.02\pm0.04$ & 25.8/23\\\cline{2-5}

                       \hline
                       
       4C+74.26 & 0.5--3 & $3.22\pm0.30$ & $0.48\pm0.17$ & 12.0/14 \\ \cline{2-5}
       			&  0.5--0.8    & $0.05\pm0.03$ & $0.04\pm0.02$ & 25.6/14\\
                        &  0.8--1    & $0.06\pm0.03$ & $0.13\pm0.03$ & 16.6/14\\
                        &  1--1.2   & $0.20\pm0.04$ & $0.10\pm0.04$ & 14.8/14\\
                        &  1.2--1.4    & $0.32\pm0.03$ & $0.04\pm0.03$ & 8.0/14\\
                        &  1.4--1.7    & $0.40\pm0.07$ & $0.08\pm0.06$ & 21.5/14\\
                        &  1.7--2    & $0.19\pm0.05$ & $0.11\pm0.05$ & 26.2/14\\
                        &  2--3    & $0.50\pm0.05$ & $0.04\pm0.05$ & 11.5/14\\\cline{2-5}
                 	  & 15--45 & $0.13\pm0.06$& $-0.01\pm0.05$ & 21.5/14\\\cline{2-5}

                        \hline
                        
    MR2251-178 & 0.5--3 & $1.39\pm0.22$ & $0.83\pm0.31$ & 31.4/26 \\ \cline{2-5}
    			&  0.5--0.8    & $0.08\pm0.03$ & $0.07\pm0.05$ & 19.8/26\\
                        &  0.8--1    & $0.07\pm0.06$ & $0.17\pm0.09$ & 42.9/26\\
                        &  1--1.2   & $0.16\pm0.06$ & $0.15\pm0.09$ & 27.7/26\\
                        &  1.2--1.4    & $0.22\pm0.07$ & $0.11\pm0.10$ & 30.2/26\\
                        &  1.4--1.7    & $0.36\pm0.05$ & $0.05\pm0.08$ & 12.7/26\\
                        &  1.7--2    & $0.27\pm0.06$ & $-0.01\pm0.09$ & 27.4/26\\
                        &  2--3    & $0.26\pm0.10$ & $0.30\pm0.14$ & 40.8/26\\

      \hline\hline

  \end{tabular}
 \end{center} 
 {\small
         \footnotemark[$*$] Errors refer to $1\sigma$ confidence limits.}

\end{table*}
\renewcommand{\arraystretch}{1.0}


\begin{table*}[t]
 \caption{Parameters, with 90\%-confidence errors, obtained by fitting the SSE spectra of the five AGNs in figure 7.}
 \label{all_parametar}
 \small
 \begin{center}
  \begin{tabular}{ccccccc}
   \hline\hline
  Model  & Parameter  & Fairall 9 & MCG-2-58-22 & 3C382 &4C+74.26&MR2251-178 \\

   \hline
   \texttt{powerlaw}     & $\Gamma$%
                       & $2.89^{+1.29}_{-1.14}$
                       & $2.92^{+1.00}_{-0.98}$
                       &$3.16^{+1.16}_{-0.93}$
                       &$3.24^{+0.98}_{-0.91}$
                       &$2.93^{+1.75}_{-2.05}$\\

                    & $N_\mathrm{PL}^{*}$%
                       & $1.92^{+0.71}_{-0.73}$
                       & $1.41^{+0.60}_{-0.61}$
                       &$2.61^{+0.74}_{-0.75}$
                       &$3.20 \pm 0.92$
                       &$3.13^{+1.55}_{-1.80}$\\[1.5ex]

 	&$\chi^{2}$/d.o.f.    & 2.28/5 & 4.09/5 & 6.88/5 & 7.36/5 & 4.50/5 \\
 
 \hline
 
  \texttt{diskbb} & $T_{\rm disk}$ (keV)
   		&$0.35^{+0.42}_{-0.18}$
		&$0.34^{+0.23}_{-0.11}$
		&$0.33^{+0.21}_{-0.13}$
		&$0.26^{+0.20}_{-0.08}$
		&$0.27^{+0.27}_{-0.11}$\\
  
   		& $N_{\rm disk}^{\dagger}$
   		&$21.7 \pm 8.2$
		&$19.5 \pm 7.3$
		&$35.9 \pm 10.3$
		&$157.1 \pm 44.7$
		&$140.1 \pm 69.4$\\[1.5ex]
		
 	&$\chi^{2}$/d.o.f.   & 3.23/5 & 2.74/5 & 6.97/5 & 7.91/5 & 4.88/5 \\

\hline

  \texttt{apec} & $T_{\rm e}$ (keV)
   		&$1.2^{+4.8}_{-0.7}$
		&$1.1^{+0.5}_{-0.3}$
		&$1.7^{+1.0}_{-0.4}$
		&$1.3^{+0.4}_{-0.6}$
		&$1.3^{+1.3}_{-0.8}$\\
		
  		 & $N_{\rm apec}^{\ddagger}$
   		&$3.29^{+2.72}_{-1.73}$
		&$2.78^{+1.35}_{-1.07}$
		&$7.11^{+2.29}_{-2.76}$
		&$7.37^{+3.24}_{-2.97}$
		&$7.14^{+5.43}_{-4.25}$\\[1.5ex]
		
 	&$\chi^{2}$/d.o.f.   & 3.49/5 & 0.25/5 & 7.29/5 & 4.48/5 & 4.65/5 \\

\hline

 \texttt{kdblur * reflionx} & $\xi^{\S}$
   		&$50.0^{+4.9}_{-16.6}$
		&$50.1^{+5.9}_{-17.5}$
		&$60.3^{+9.0}_{-7.8}$
		&$36.3^{+9.1}_{-7.3}$
		&$22.8^{+4.1}_{-6.1}$\\
		
  		 & $N_{\rm reflionx}^{||}$
   		&$0.62 \pm 0.23$
		&$0.42 \pm 0.17$
		&$0.61 \pm 0.18$
		&$1.97 \pm 0.47$
		&$5.09 \pm 1.90$\\[1.5ex]
		
 	&$\chi^{2}$/d.o.f.   & 2.37/5 & 7.38/4 & 7.29/4 & 4.48/4 & 4.65/4 \\

\hline

  \texttt{comptt} & $T_{\rm e}$ (keV)
   		&$0.80^{+1.23}_{-0.38}$
		&$0.26^{+0.11}_{-0.07}$
		&$0.46^{+0.27}_{-0.17}$
		&$0.77^{+0.60}_{-0.28}$
		&$0.86^{+0.09}_{-0.13}$\\
		
  		 & $\tau$
   		&$16.2^{+11.1}_{-5.6}$
		&$94.1^{+28.1}_{-21.0}$
		&$23.9^{+17.3}_{-8.3}$
		&$14.9^{+6.9}_{-4.0}$
		&$15.2^{+179.2}_{-10.2}$\\
  		
		& $N_{\rm comp}^{\#}$
   		&$0.21 \pm 0.08$
		&$0.03 \pm 0.01$
		&$0.41 \pm 0.12 $
		&$0.83 \pm 0.24$
		&$0.42 \pm 0.14$\\[1.5ex]
		
	& $\chi^{2}$/d.o.f.   & 2.39/4 & 2.59/4 & 6.63/4 & 7.35/4 & 4.57/4 \\
\hline
 	&$L^{**}$
   		&$1.94^{+0.39}_{-0.45}$
		&$0.42 \pm 0.12$
		&$3.66^{+0.72}_{-0.50}$
		&$10.2^{+1.6}_{-1.9}$
		&$6.47^{+2.13}_{-1.34}$\\

 \hline\hline
      
  \end{tabular}
 \end{center}
   
         {\small
         \footnotemark[$*$] The cutoff power-law normalization at 1 keV, in units of $10^{-3}$~photons~keV$^{-1}$~cm$^{-2}$~s$^{-1}$~at 1 keV. \\
          \footnotemark[$\dagger$] The \texttt{diskbb} normalization, in $((R_{\rm in}/$km$)/(D/10~$kpc$))^2 \cos \theta$, where $R_{\rm in}$ is the inner disk radius, $D$ the distance to the source, and $\theta$ the angle of the disk. \\
           \footnotemark[$\ddagger$] The \texttt{apec} normalization, in units of $(10^{-17}/4 \pi (D_{\rm A} (1+z))^{2}) \int n_{\rm e} n_{\rm H} dV$, where $D_{\rm A}$ is the angular size distance to the source (cm), $n_{\rm e}$ and $n_{\rm H}$ are the electron and H densities (cm$^{-3}$).\\
            \footnotemark[$\S$] The ionization parameter, in units of erg cm$^{-2}$ s$^{-1}$.\\
            \footnotemark[$||$]  The \texttt{reflionx} normalization, in units of $10^{-5}$~photons~keV$^{-1}$~cm$^{-2}$~s$^{-1}$~at 1keV. \\
            \footnotemark[$\#$] The \texttt{comptt} normalization, in units of ~photons~keV$^{-1}$~cm$^{-2}$~s$^{-1}$~at 1 keV.  \\
            \footnotemark[$**$] The 0.5--3 keV luminosity of the extracted SSEs, in units of $10^{43}$~erg s$^{-1}$. } 
   
\end{table*}

While a soft excess component was usually identified as excess signals above the 
PL determined by higher-energy spectral data, Noda et al. (2011b) formulated 
an independent way of doing this. Below, we follow their C3PO method.  
For this purpose, 0.5--3 keV (filled circles), 3--10 keV (open circles), and 15--45 keV (open squares, from HXD-PIN) 
light curves of the five AGNs are shown in figure 1, after subtracting the NXB. 
All of the XIS FI light curves exhibit more than 10\% variations, 
with apparently tight correlations between the two XIS bands. 
To quantify this behavior, 
we produced Count-Count Plots (CCPs), 
in which abscissa is the 3--10 keV band XIS FI count rates, 
and ordinate is  those in the 0.5--3 keV.  
The results, shown in panel (a) of figure 2 to figure 7, indeed indicate that 
the 0.5--3 keV count rate, to be denoted $y$, varies linearly with that in 3--10 keV, $x$. 

For further quantification, 
we fitted the data in each CCP with one straight line, 
expressed as 

\begin{eqnarray}
y = a x + b,
\label{eq:ccpfit12}
\end{eqnarray}
in which the two parameters, $a$ and $b$, were both left free.
The fit goodness was evaluated in terms of chi-square statistics using errors expressed as 
$\sigma = \sqrt{\sigma_{y}^2 + (a \sigma_{x})^2}$, 
where $\sigma_{x}$ and $\sigma_{y}$ are 1-sigma statistical errors 
associated with $x$ and $y$, respectively.

As shown in table 2 and indicated in panel (a) of figures 2--6, 
the data distribution in each CCP has been reproduced successfully with equation (1). 
Therefore, the main variable component 
is considered to have varied without any change in its shape.
In addition,  the offset $b$ is significantly positive in all cases (2.4--2.8$\sigma$),  
indicating that some signals should remain in the 0.5--3 keV bands even when the  intensity
of the variable component becomes zero.
This condition is the same as found with Mrk 509 by Noda et al. (2011b), 
and allows us to apply the C3PO method to the present five AGNs.

We divided the 0.5--3 keV band of the individual AGNs into seven finer bands,
0.5--0.8 keV, 0.8--1 keV, 1--1.2 keV, 1.2--1.4 keV, 
1.4--1.7 keV, 1.7--2 keV, and 2--3 keV.
Then, for each AGN, we created seven additional CCPs, where abscissa is again 
the 3--10 keV count rate while ordinate is those in the seven finer soft X-ray bands. 
The obtained CCPs are presented in panels (b) to (h) of figure 2 to 6. 

To examine the data distribution in each CCP representing a finer energy band, 
the fitting with one straight line expressed by  equation (1) was again performed. 
The obtained parameter values are summarized in table 2. 
Thus, the fit with equation (1) was successful in most of the 35 CCPs (the seven bands of the five objects). 
In all objects, furthermore,  the CCPs generally exhibit positive offsets ($b>0$), 
particularly in lower energy bands. 
These offsets, divided by the corresponding energy intervals, 
define a spectrum that represents the non-varying soft X-ray signals of each AGN. 
In figure 7, we show these spectra in purple, after further normalizing to unabsorbed PL models 
with a photon index $\Gamma=2$ of which the normalizations were chosen to approximately reproduce the 
observed time-averaged spectra of the corresponding objects.

To quantify these stable soft emission (hereafter SSE) spectra, we fitted them with five representative models; 
power law (\texttt{powerlaw}), multi-color disk (\texttt{diskbb}; Mitsuda et al. 1984) emission, 
thin-thermal plasma radiation (\texttt{apec}; Smith et al. 2001),  
relativistically smeared and ionized reflection (\texttt{kdblur * reflionx}; Laor 1991, Ross \& Fabian 2005),  
and thermal Comptonization (\texttt{comptt}; Titarchuk 1994). 
The fits all include the Galactic line-of-sight absorption, which is modeled by \texttt{wabs} (Morrison and McCammon 1983) 
with the $N_{\rm H}$ values given in table 1. 
In fitting with the \texttt{apec} model, the abundance parameter was fixed at 0.5 solar,  
while the temperature and the normalization were left free. 
In the \texttt{kdblur * reflionx} model, the photon index of the primary continuum for reflection was fixed at 2, 
and the inner and outer disk radii at 1.24~$R_{\rm g}$ and 400~$R_{\rm g}$, respectively. 
The emissivity index was also fixed at 4. The other parameters,  
the inner radius, the ionization parameters, and the normalization, were all left free. 
In the \texttt{comptt} fit, the seed photon temperature was fixed at 0.02 keV, while the other parameters were left free. 
The redshifts in those models were fixed at the values given in table 1. 
The obtained fit parameters are summarized in table 3, 
which  also gives the 0.5--3 keV luminosity of this component calculated by the \texttt{apec} model.  
Thus, the five SSE spectra can be reproduced successfully by any of the employed five models. 
They are characterized by a PL photon index of $\sim 3$, or a disk temperature of $\sim 0.3$ keV, 
or an \texttt{apec} temperature of $\sim 1.2$ keV. 
If employing the relativistically blurred and ionized reflection model, 
the ionization parameter became $\sim 50$~erg cm$^{-2}$ s$^{-1}$, with the inner disk radius fixed at 1.24~$R_{\rm g}$. 
When employing the thermal Comptonization model with a seed photon temperature of $0.02$ keV, 
which can be calculated from  the mass of a typical central BH of $\sim10^8 M_{\odot}$ 
and an Eddington ratio of $\sim10\%$, 
the coronal temperature becomes 0.3--1 keV and the optical depth $\gtrsim 15$. 
Below, we do not consider the \texttt{diskbb} modeling, 
since its temperature is too high as already pointed out previously (section 1). 
The PL modeling may not be considered either, because the indicated steep PL 
does not accept easy interpretations. In other words, 
we retain the \texttt{apec}, \texttt{kdblur * reflionx}, and \texttt{comptt} modelings. 

\begin{figure*}[t]
 \begin{center}
   \FigureFile(155mm,155mm)
    {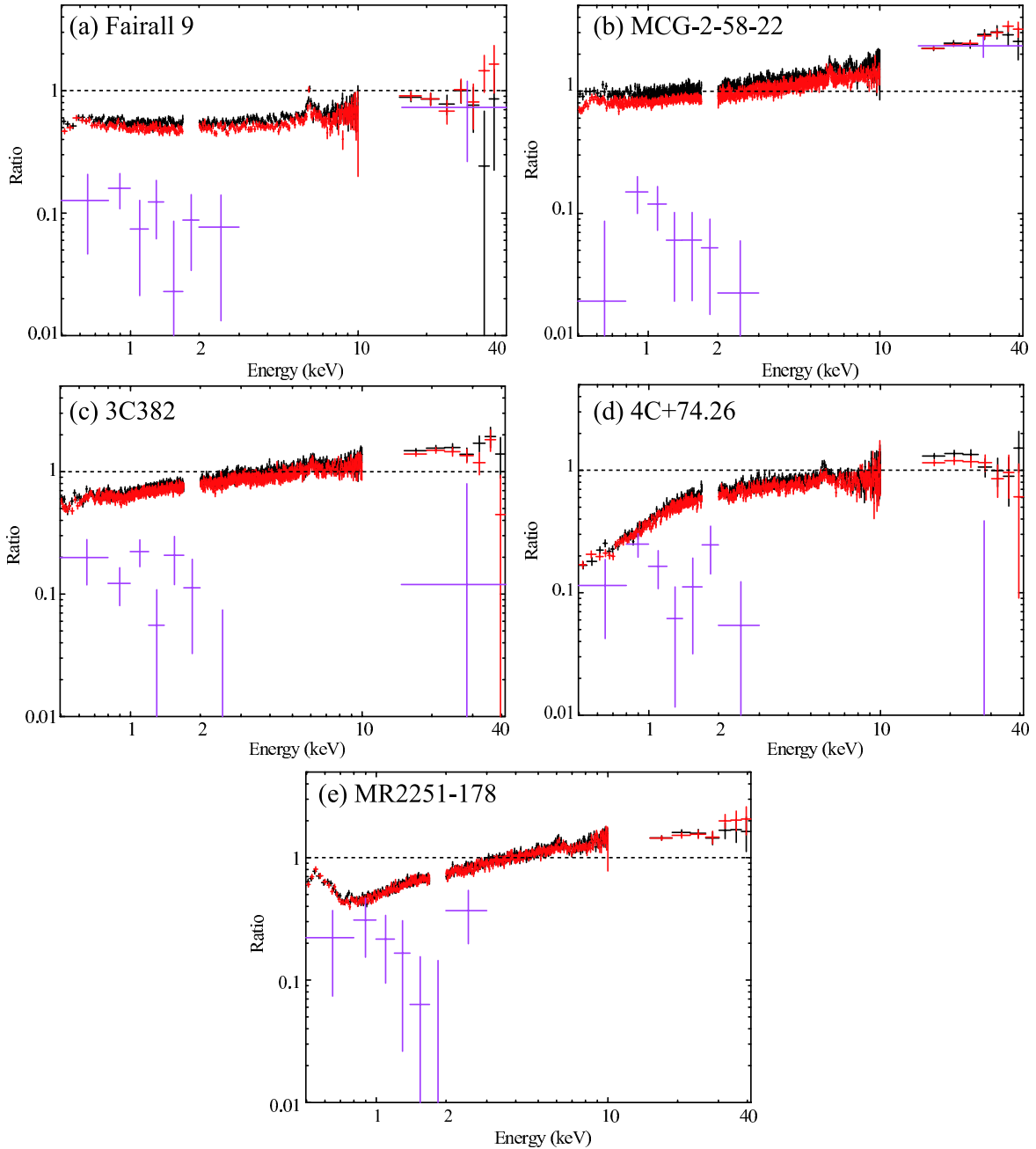}
 \end{center}
 \caption
{The High (black) and Low (red) spectra
of the present sample AGNs, shown as a ratio to  a common unabsorbed $\Gamma = 2$ PL. 
Purple shows the SSE and SHE spectra of each object determined by $b$ of table 2, presented after divided by 
the same PL as used to normalize the High and Low spectra.}

\label{fig:lcs}
\end{figure*}

\begin{figure*}[t]
 \begin{center}
   \FigureFile(155mm,155mm)
    {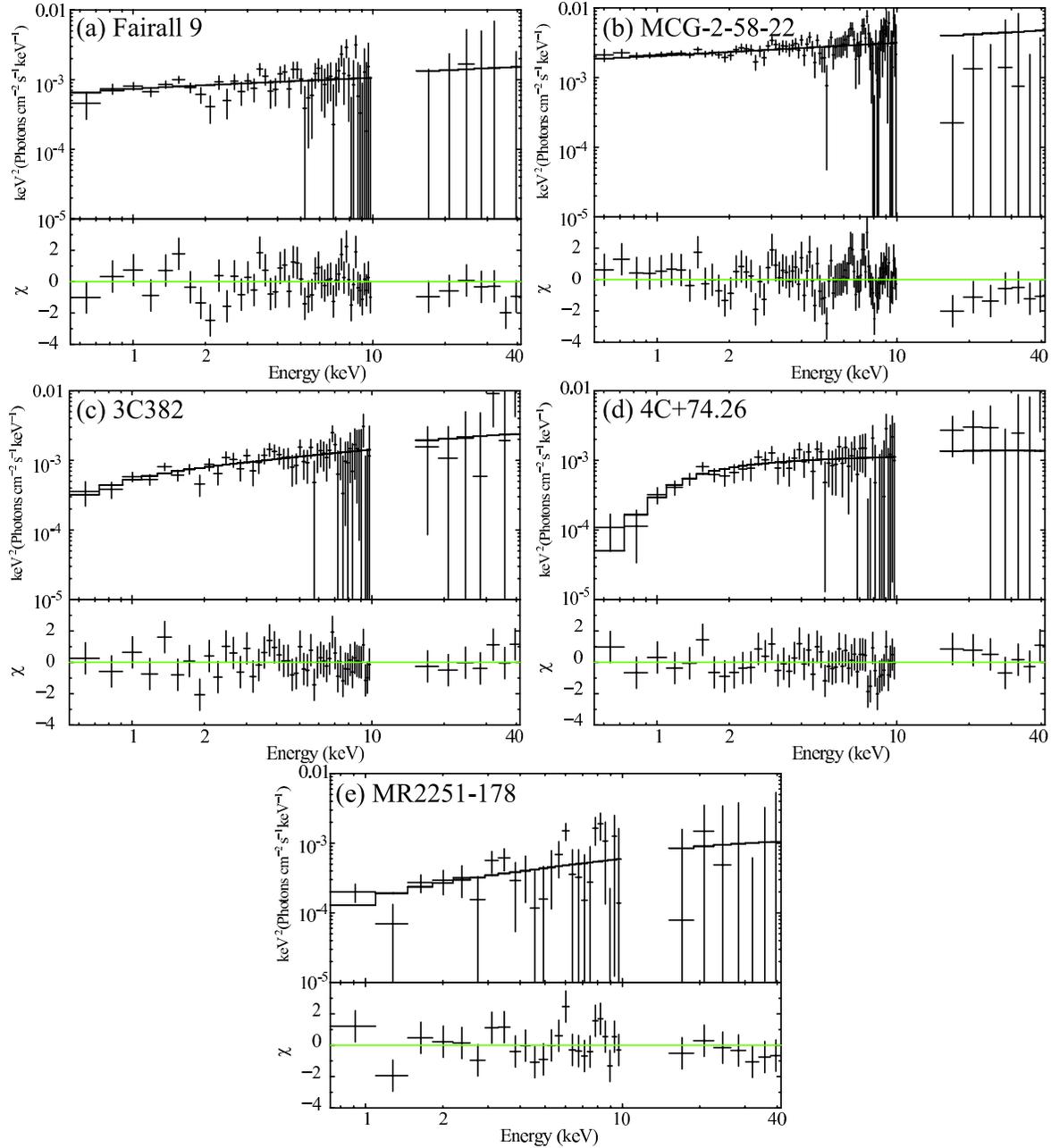}
 \end{center}
 \caption
{Difference spectra of the AGNs in $\nu F_{\nu}$ forms, obtained by subtracting Low-phase 
from High-phase spectra, and fitted by \texttt{wabs * cutoffpl}.
The lower panels show residuals from the best-fit model. }

\label{fig:lcs}
\end{figure*}


\begin{table*}[t]
 \caption{The parameters obtained in the fits to 0.5--45 keV difference 
 spectra of the five AGNs.}
 \label{all_parametar}
 \small
 \begin{center}
  \begin{tabular}{ccccccc}
   \hline\hline
    &  & Fairall 9 & MCG-2-58-22 & 3C382 &4C+74.26&MR2251-178 \\

   \hline
  \texttt{wabs} & $N_{\rm H}^{*}$
   		&$<0.10$
		&$<0.03$
		&$0.08^{+0.13}_{-0.08}$
		&$0.43^{+0.27}_{-0.21}$
		&$<0.50$\\[1.5ex]

powerlaw     & $\Gamma$%
                       & $1.84^{+0.19}_{-0.13}$
                       & $1.80^{+0.06}_{-0.05}$
                       &$1.63^{+0.18}_{-0.16}$
                       &$1.86^{+0.28}_{-0.24}$
                       &$1.40^{+0.43}_{-0.32}$\\

                    & $N_\mathrm{PL}^{\dagger}$%
                       & $0.75^{+0.18}_{-0.09}$
                       & $0.32^{+0.19}_{-0.11}$
                       &$0.64^{+0.17}_{-0.13}$
                       &$0.85^{+0.40}_{-0.25}$
                       &$0.017^{+0.018}_{-0.006}$\\ \hline

   $\chi^{2}$/d.o.f. &   & 58.13/55 & 100.15/90 & 35.97/55 & 35.09/55 & 28.43/32 \\
      \hline\hline
      
  \end{tabular}
 \end{center}
   
         {\small
         \footnotemark[$*$] Equivalent hydrogen column density in  $10^{22}$ cm$^{-2}$.  \\
         \footnotemark[$\dagger$] The cutoff power-law normalization at 1 keV, in units of $10^{-3}$~photons~keV$^{-1}$~cm$^{-2}$~s$^{-1}$. } 
   
\end{table*}

In an attempt to extend our analysis to the HXD data,
we additionally show, in figure 1, 15--45 keV light curves of the five AGNs obtained with HXD-PIN.
The CCPs of these bands against the 3--10 keV XIS counts
are shown in figure 2--6(i).
The results of fitting these CCPs with equation (1) are given in table 2, 
except for MR2251-178 in which the data points have too large errors and 
the variation is too small to derive meaningful constraints on the parameters $a$ and $b$. 
Thus, the hard-band CCPs of Fairall 9 and MCG-2-58-22 suggest 
statistically significant positive offsets (though with large errors),  
so we divided them by the energy range of 20 keV, 
and show the results in figure 7 as ratios to the same $\Gamma=2$ PL. 
Hereafter, these hard-band components derived with the HXD vs. XIS C3PO method 
are called stable hard emissions (SHEs).

\subsection{High, low, and defference spectra}
\label{sec:HighLow}

In figure 7, time-averaged and background-subtracted spectra of our AGNs are also shown, 
normalized to the same (including the normalization) $\Gamma=2$ PL 
as used to display the SSE and SHE spectra. 
However, instead of summing all photons of each AGN into a single spectrum, 
we accumulated them over two phases, hereafter called High and Low phases, 
when the 3--10 keV source intensity is higher (black) 
and lower (red) than the average (dotted line in figure 1), respectively. 

To examine shapes of the main variable spectral components, presumably PL-like,  
we subtracted Low from High spectra, and show the derived difference spectra of the five 
AGNs in figure 8(a--e). 
Each difference spectrum was fitted successfully by an absorbed cutoff PL model (\texttt{wabs * cutoffpl} in XSPEC), 
in which the photon index and normalization, as well as the absorbing column $N_{\rm H}$, 
were left free, while the cutoff energy was fixed at 200 keV. 
Therefore, in these AGNs, the varying signals can be expressed within errors by an absorbed PL 
(with an assumed cutoff), in agreement with some previous studies (e.g., Schmoll et al. 2009; 
Rivers et al. 2011; Sambruna et al. 2011; Larsson et al. 2008; and Gofford et al. 2011). 
This also explains the good linearity found with our CCPs. 
The obtained parameters are summarized in table 4. 
The weakly-absorbed AGNs, Fairall 9, MCG-2-58-22, and 3C382, 
indeed exhibit low absorptions, 
which mostly agree, within their $90\%$ errors,  with the Galactic line-of-sight column densities
reported by Dickey and Lockman (1990) and quoted in table 1. 
Conversely, the strongly-absorbed AGN, 4C+74.26, exhibits significantly higher absorption 
than the Galactic value (table 1).

\subsection{Time-averaged spectrum  analysis}
\label{sec:spectrum}

\begin{figure*}[t]
 \begin{center}
   \FigureFile(155mm,155mm)
    {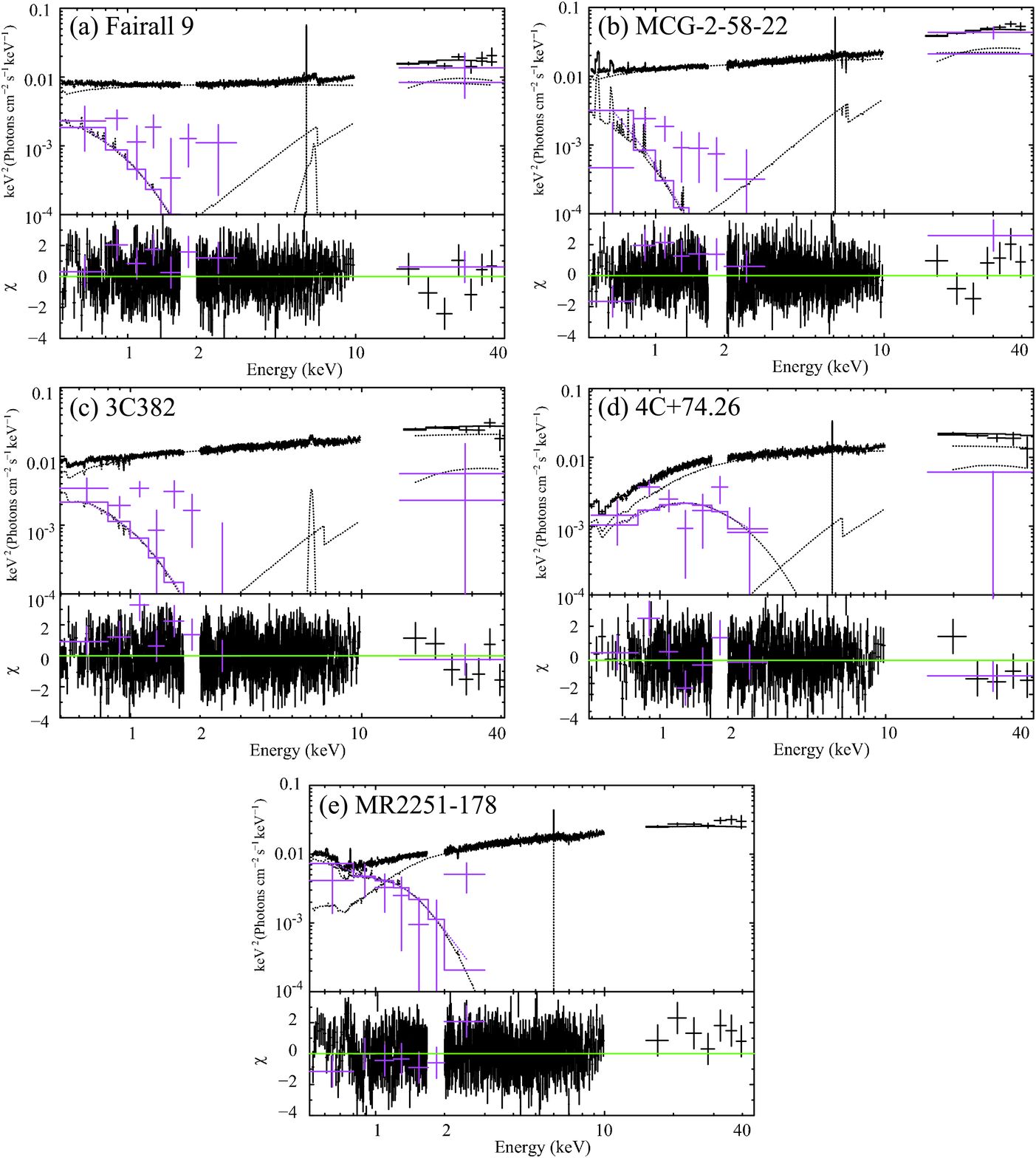}
 \end{center}
 \caption
{Fits to the time-averaged spectra of the five AGN, with models of 
\texttt{wabs * (cutoffpl + pexrav + zgauss + laor + apec)} for Fairall 9, 
\texttt{wabs * zxipcf * (cutoffpl + pexrav + zgauss + 3 $\times$ zgauss + apec)} for MR2251-178, 
and  \texttt{wabs * zxipcf * (cutoffpl + pexrav + zgauss + apec)} for the others, respectively.
The SSE and SHE components are  also shown in purple.}

\label{fig:lcs}
\end{figure*}

In subsections 4.1 and 4.2, 
we followed the C3PO method of Noda et al. (2011b), 
and utilized time variations to extract the three distinct spectral components; 
the SSE and SHE components (purple in figure 7), 
and the variable (absorbed) PL component (figure 8). 
Then, can we arrive at the same spectral decomposition by analyzing the time-averaged 
(i.e., High-phase plus Low-phase) wide-band spectra of our sample objects? 
To see this, we fitted these spectra with a model of  
\texttt{wabs * zxipcf * (cutoffpl + pexrav + zgauss)}. 
Here, \texttt{pexrav} represents a reflection continuum from neutral matters (Magdziarz \& Zdziarski 1995), 
and \texttt{zgauss} an Fe {\emissiontype I} K$\alpha$ line. 
\texttt{zxipcf} (Reeves et al. 2008) represents the absorption by an ionized (or ``warm'') absorber, 
with its equivalent hydrogen column density, the ionization parameter $\xi$, 
and covering fraction left as free parameters.  
This follows Weaver et al. (1995), Sambruna et al. (2011),  Kaspi et al. (2004),  and 
Gofford et al. (2011), who reported the presence of warm absorbers in MCG-2-58-22, 
3C382, 4C+74.26, and MR2251-178, respectively. 
For Fairall 9, a broad iron line component  \texttt{laor} (Laor 1991) was added, 
while the \texttt{zxipcf} factor was excluded 
because this is considered to be a bare-nucleus object 
(e.g., Schmoll et al. 2008; Patrick et al. 2011). 
In addition, for MR2251-178 which was reported to exhibit absorption 
by outflows (Gofford et al. 2011), 
we added three gaussians with free $\sigma$ and negative normalization, 
to represent Fe$_{\rm I-XVI}$ M-shell unresolved transition forest  
(at a center energy of $E_{\rm c}=0.77$~keV), as well as those from Fe$_{\rm XXIV}$ L-shell ($E_{\rm c}=1.29$~keV) and 
Fe$_{\rm XXV-XXVI}$ K-shell ($E_{\rm c} =7.57$~keV),  
and fitted with a model of \texttt{wabs * zxipcf * (cutoffpl + pexrav + zgauss + 3 $\times$ zgauss)}.
The cutoff energies were fixed at 200~keV, except for MR2251-178 in which it was 
fixed at 100~keV after Gofford et al. (2011). 
The redshifts in these models were all fixed at the values given in table 1. 
As a result, 
we obtained $\chi^2/$d.o.f. of  1105.61/658 for Fairall 9, 1099.49/844 for MCG-2-58-22, 
1070.05/846 for 3C382, 666.18/605 for 4C+74.26, and 1581.01/871 for MR2251-178:  
all the five AGNs, except for 4C+74.26, rule out the simple PL-based modeling. 
These fit failures in fact arise mainly due to data excess above the model in soft X-rays, 
which does not vanish even when the absorbing column density is left free to vary. 
Through this ``static" spectral analysis, 
we have thus confirmed the presence of a soft excess component, in at least four of our sample objects. 

\begin{table*}[t]
 \caption{The results of  fitting the time-averaged spectra of the five AGNs, 
 wherein the soft excess is represented by an \texttt{apec} model. }
 \label{all_parametar}
 \small
 \begin{center}
  \begin{tabular}{ccccccc}
   \hline\hline
   Model & Parameter & Fairall 9 & MCG-2-58-22 & 3C382 &4C+74.26 &MR2251-175 \\

   \hline
  \texttt{wabs} & $N_{\rm H}^{*}$
   		& $0.032$(fix)
		&$0.035$(fix)
		& $0.074$(fix)
		&$0.21^{+0.03}_{-0.01}$
		&$0.028$(fix)\\

 \texttt{zxipcf} & $N_{\rm H}^{*}$
   		           & --
		            &$7.27^{+5.52}_{-4.41}$
		            &$0.07^{+0.19}_{-0.02}$
		            &$1.36^{+0.28}_{-0.05}$
		            &$1.07 \pm 0.02$\\
		            
		             &log$\xi$
		             &--
		             &$4.35^{+1.63}_{-0.57}$
		             &$2.55^{+0.53}_{-0.16}$
		             &$0.53^{+0.11}_{-0.09}$
		             &$0.40^{+0.01}_{-0.03}$ \\
		             
		             &Cvr frac.
		             &--
		             &$0.58^{+0.12}_{-0.38}$
		             &$>0.62$
		             &$0.61^{+0.04}_{-0.05}$
		             &$0.86^{+0.01}_{-0.02}$\\[1.5ex]

  powerlaw     & $\Gamma$%
                       & $1.99^{+0.02}_{-0.03}$
                       & $1.85 \pm 0.01$
                       &$1.82 \pm 0.01$
                       &$1.92^{+0.08}_{-0.08}$
                       &$1.72 \pm 0.01$\\

                    & $N_\mathrm{PL}^{\dagger}$%
                       & $0.78 \pm 0.01$
                       & $1.34^{+0.02}_{-0.01}$
                       &$1.11 \pm 0.01$
                       &$1.10^{+0.06}_{-0.08}$
                       &$1.15^{+0.06}_{-0.01}$\\[1.5ex]
                       
    \texttt{pexrav}  & $f_\mathrm{ref}$%
                          &$1.9^{+0.4}_{-0.3}$
                          &$1.8 \pm 0.1$
                          &$0.5^{+0.1}_{-0.2}$
                          &$0.9^{+0.4}_{-0.5}$
                          &$<0.2$\\[1.5ex]

 Fe~I~K$\alpha$   & $E_\mathrm{c}^{\ddagger}$

                        & $6.37 \pm 0.02$
                        & $6.25 \pm 0.02$
                        &$6.41 \pm 0.03$
                        &$6.40 \pm 0.04$
                        &$6.38^{+0.04}_{-0.03}$\\
                                
                                   & $\sigma$~(keV)
                               &$10^{-4}$(fix)
                               &$10^{-4}$(fix)
                           	&$0.099 \pm 0.032$
			&$10^{-4}$(fix)
			&$10^{-4}$(fix)\\

                   & $N_\mathrm{Fe}^{\S}$ %
                        & $2.30^{+0.21}_{-0.22}$
                        & $1.84 \pm 0.32$
                        &$2.25^{+0.44}_{-0.43}$
                        &$1.42^{+0.35}_{-0.34}$
                        &$1.15^{+0.30}_{-0.29}$\\                    

                    & $EW$ (eV) %
                        & $77^{+40}_{-62}$
                        & $33^{+54}_{-32}$
                        &$49^{+27}_{-15}$
                        &$30^{+32}_{-12}$
                        &$21^{+8}_{-6}$\\[1.5ex]

\texttt{laor} & $E_\mathrm{c}^{||}$
		&$6.37 \pm 0.07$
		&--
		&--
		&--
		&--\\
		
		&$R_{\rm in}$~($R_{\rm g}$)
		&$16.7^{+28.0}_{-7.8}$
		&--
		&--
		&--
		&--\\
		
		& $N_\mathrm{Fe}^{\#}$
		&$1.67^{+0.57}_{-0.50}$
		&--
		&--
		&--
		&--\\
		
		& $EW$ (eV) %
                        & $204^{+56}_{-20}$
                        &--
                        &--
                        &--
                        &--\\[1.5ex]

  \texttt{apec}  & $kT$~(keV)
		 &$0.32^{+0.02}_{-0.04}$
		 &$0.18^{+0.02}_{-0.03}$
		 &$0.25^{+0.03}_{-0.04}$
		 &$0.68^{+0.08}_{-0.07}$
		 &$0.31 \pm 0.01$\\
                        
                   &$A$~(Z$_{\odot}$)     
                        &$<0.005$
                        &$<0.007$
                        &$<0.004$
                        &$<0.005$
                        &$0.002 \pm 0.001$\\      
                        
                     &$N_{\rm apec}^{**}$   
                        &$0.52^{+0.15}_{-0.12}$
                        &$1.41^{+0.64}_{-0.42}$
                        &$1.09^{+0.45}_{-0.24}$
                        &$0.73^{+0.09}_{-0.12}$
                        &$8.06^{+1.47}_{-0.87}$\\   \hline

   $\chi^{2}$/d.o.f.&   & 728.62/651  & 846.70/832 & 883.80/842 &612.45/601 & 983.20/865 \\
      \hline\hline
      
  \end{tabular}
 \end{center}
   
     	{\small
	\footnotemark[$*$] Equivalent hydrogen column density in  $10^{22}$ cm$^{-2}$ for the Galactic or the intrinsic line-of-sight absorption.  \\
         \footnotemark[$\dagger$] The power-law normalization at 1 keV, in units of $10^{-2}$~photons~keV$^{-1}$~cm$^{-2}$~s$^{-1}$~at 1 keV. \\
         \footnotemark[$\ddagger$] Center energy in keV in the rest frame.\\
         \footnotemark[$\S$] The Gaussian normalization in units of 
         $10^{-5}$~photons~cm$^{-2}$~s$^{-1}$.\\
         \footnotemark[$||$] Center energy in units of keV. \\
         \footnotemark[$\#$] The \texttt{loar} Normalizatiation in units of 
         $10^{-5}$~photons~cm$^{-2}$~s$^{-1}$. \\
           \footnotemark[$**$] The \texttt{apec} normalization, in units of $(10^{-15}/4 \pi (D_{\rm A} (1+z))^{2}) \int n_{\rm e} n_{\rm H} dV$, where $D_{\rm A}$ is the angular size distance to the source (cm), $n_{\rm e}$ and $n_{\rm H}$ are the electron and H densities (cm$^{-3}$). 
         }

\end{table*}


\begin{figure*}[t]
 \begin{center}
   \FigureFile(155mm,155mm)
    {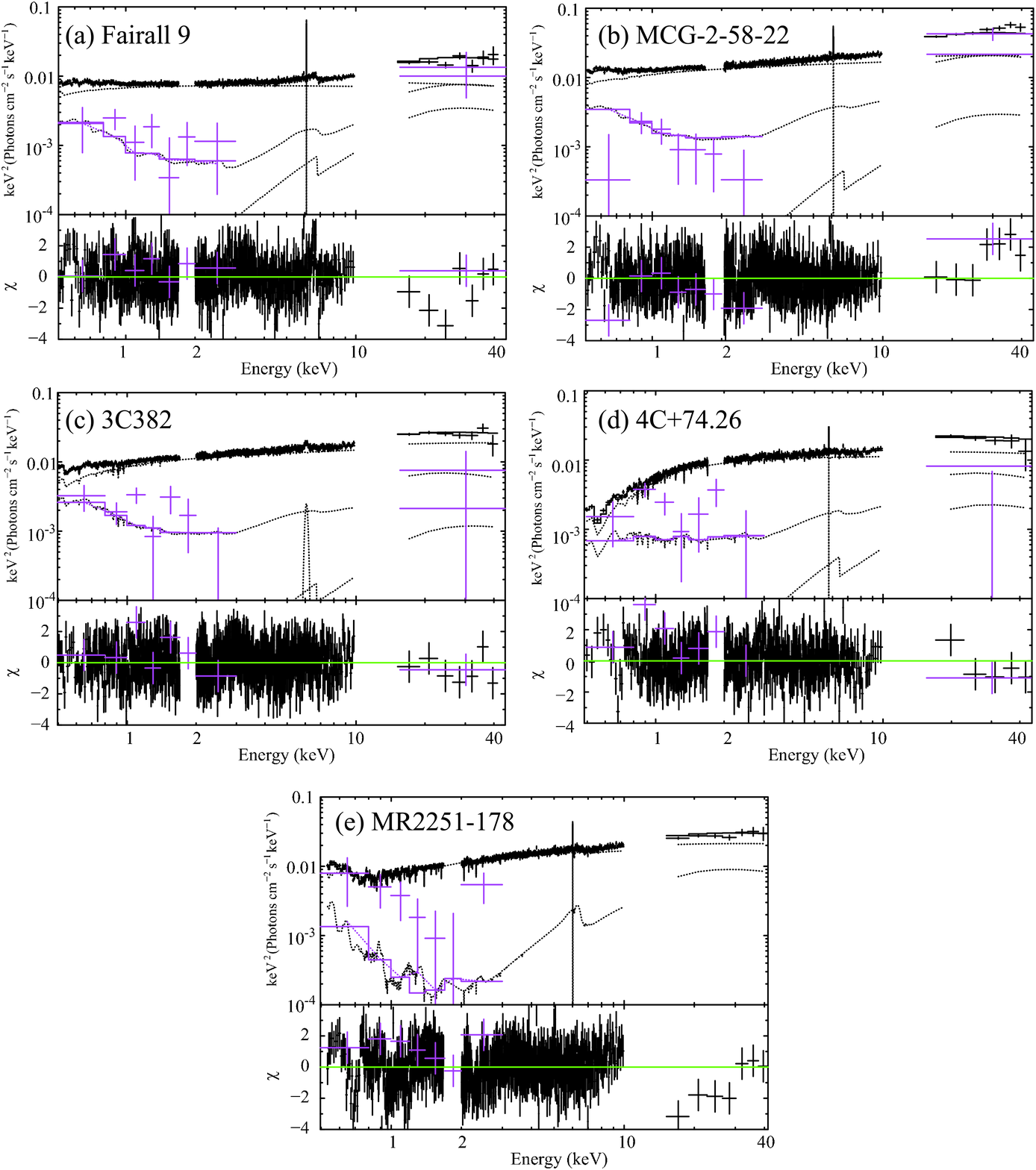}
 \end{center}
 \caption
{Same as figure 9, with the \texttt{apec} model replaced by a \texttt{kdblur * reflionx} model. 
The spectrum of Fairall 9 was fitted with \texttt{wabs * (cutoffpl + pexrav + zgauss + kdblur * reflionx)}, 
and that of MR2251-178 was with 
\texttt{wabs * zxipcf * (cutoffpl + pexrav + zgauss + 3 $\times$ zgauss + kdblur * reflionx)}, 
while those of the others were with \texttt{wabs * zxipcf * (cutoffpl + pexrav + zgauss + kdblur * reflionx)}. }

\label{fig:lcs}
\end{figure*}


\begin{table*}[t]
 \caption{The same as table 5, but the soft excess is represented by ionized and relativistically-blurred 
 reflection (\texttt{kdblur * reflionx})  instead of \texttt{apec}.}
 \label{all_parametar}
 \small
 \begin{center}
  \begin{tabular}{ccccccc}
   \hline\hline
   Model & Parameter & Fairall 9 & MCG-2-58-22 & 3C382 &4C+74.26 &MR2251-175 \\

   \hline
  \texttt{wabs} & $N_{\rm H}$
   		& $0.032$(fix)
		&$0.035$(fix)
		& $0.074$(fix)
		&$0.26^{+0.03}_{-0.02}$
		&$0.028$(fix)\\

 \texttt{zxipcf} & $N_{\rm H}$
   		           & --
		            &$< 0.36$
		            &$< 0.12$
		            &$0.24^{+1.11}_{-0.07}$
		            &$3.15 \pm 0.02$\\
		            
		             &log$\xi$
		             &--
		             &$0.73^{+0.32}_{-0.24}$
		             &$2.52 \pm 0.16$
		             &$2.26^{+0.16}_{-0.13}$
		             &$2.16^{+0.66}_{-0.58}$ \\
		             
		             &Cvr frac.
		             &--
		             &$< 0.25$
		             &$>0.43$
		             &$>0.21$
		             &$0.65^{+0.09}_{-0.15}$\\[1.5ex]

  powerlaw     & $\Gamma$%
                       & $1.99 \pm 0.03$
                       & $1.84^{+0.02}_{-0.01}$
                       &$1.82 \pm 0.02$
                       &$1.92^{+0.06}_{-0.04}$
                       &$1.83^{+0.01}_{-0.02}$\\

                    & $N_\mathrm{PL}$%
                       & $0.74 \pm 0.01$
                       & $1.22 \pm 0.02$
                       &$1.04^{+0.01}_{-0.02}$
                       &$0.99^{+0.08}_{-0.06}$
                       &$1.23^{+0.04}_{-0.02}$\\[1.5ex]
                       
   \texttt{pexrav}  & $f_\mathrm{ref}$%
                          &$0.8 \pm 0.3$
                          &$<0.3$
                          &$<0.2$
                          &$<0.5$
                          &$<0.2$\\[1.5ex]

 Fe~I~K$\alpha$   & $E_\mathrm{c}$

                        & $6.38^{+0.04}_{-0.03}$
                        & $6.39 \pm 0.03$
                        &$6.42^{+0.03}_{-0.04}$
                        &$6.39 \pm 0.04$
                        &$6.40^{+0.04}_{-0.03}$\\
                                
                                   & $\sigma$~(keV)
                               &$10^{-4}$(fix)
                               &$10^{-4}$(fix)
                           	&$0.095^{+0.025}_{-0.029}$
			&$10^{-4}$(fix)
			&$10^{-4}$(fix)\\

                   & $N_\mathrm{Fe}$ %
                        & $2.45^{+0.25}_{-0.20}$
                        & $2.01^{+0.21}_{-0.28}$
                        &$2.25^{+0.44}_{-0.43}$
                        &$1.46^{+0.31}_{-0.38}$
                        &$0.91 \pm 0.25$\\                    

                    & $EW$ (eV) %
                        & $91^{+40}_{-62}$
                        & $33^{+44}_{-31}$
                        &$52^{+27}_{-15}$
                        &$37^{+32}_{-12}$
                        &$25^{+12}_{-10}$\\[1.5ex]

\texttt{kdblur} & $q^{*}$
		&$4.48^{+2.70}_{-0.98}$
		&$5.15^{+4.19}_{-0.81}$
		&$4.87^{+3.11}_{-0.68}$
		&$3.96^{+0.89}_{-1.35}$
		&$2.45^{+2.31}_{-0.27}$\\
		
		&$R_{\rm in}$~($R_{\rm g}$)
		&$2.30^{+0.49}_{-0.16}$
		&$1.24^{+0.35}_{-0.0}$
		&$2.44^{+0.56}_{-0.22}$
		&$1.24^{+0.31}_{-0.0}$
		&$1.24^{+0.27}_{-0.0}$\\[1.5ex]

  \texttt{reflionx}  & $\xi^{\dagger}$
		 &$58.9^{+27.2}_{-31.2}$
		 &$101.8^{+53.2}_{-48.4}$
		 &$103.5^{+56.6}_{-50.8}$
		 &$94.1^{+42.9}_{-34.8}$
		 &$20.2^{+12.1}_{-5.2}$\\
                                                
                     &$N_{\rm reflionx}^{\ddagger}$   
                        &$3.11^{+0.34}_{-0.12}$
                        &$4.97^{+0.64}_{-0.42}$
                        &$2.39^{+0.45}_{-0.24}$
                        &$2.44^{+0.30}_{-0.12}$
                        &$1.28^{+0.23}_{-0.09}$\\   \hline

   $\chi^{2}$/d.o.f.&   & 785.06/651  & 945.79/832 & 886.26/842 &660.51/601 & 1145.17/865 \\
      \hline\hline
      
  \end{tabular}
 \end{center}
   
     	{\small
       \footnotemark[$*$] The emissivity index (scales as $R^{-q}$). \\
          \footnotemark[$\dagger$] The ionization parameter, in units of erg cm$^{-2}$ s$^{-1}$. \\
         \footnotemark[$\ddagger$] The \texttt{reflionx} normalization, in units of $10^{-6}$~photons~keV$^{-1}$~cm$^{-2}$~s$^{-1}$~at 1 keV.          }

\end{table*}


\begin{figure*}[t]
 \begin{center}
   \FigureFile(155mm,155mm)
    {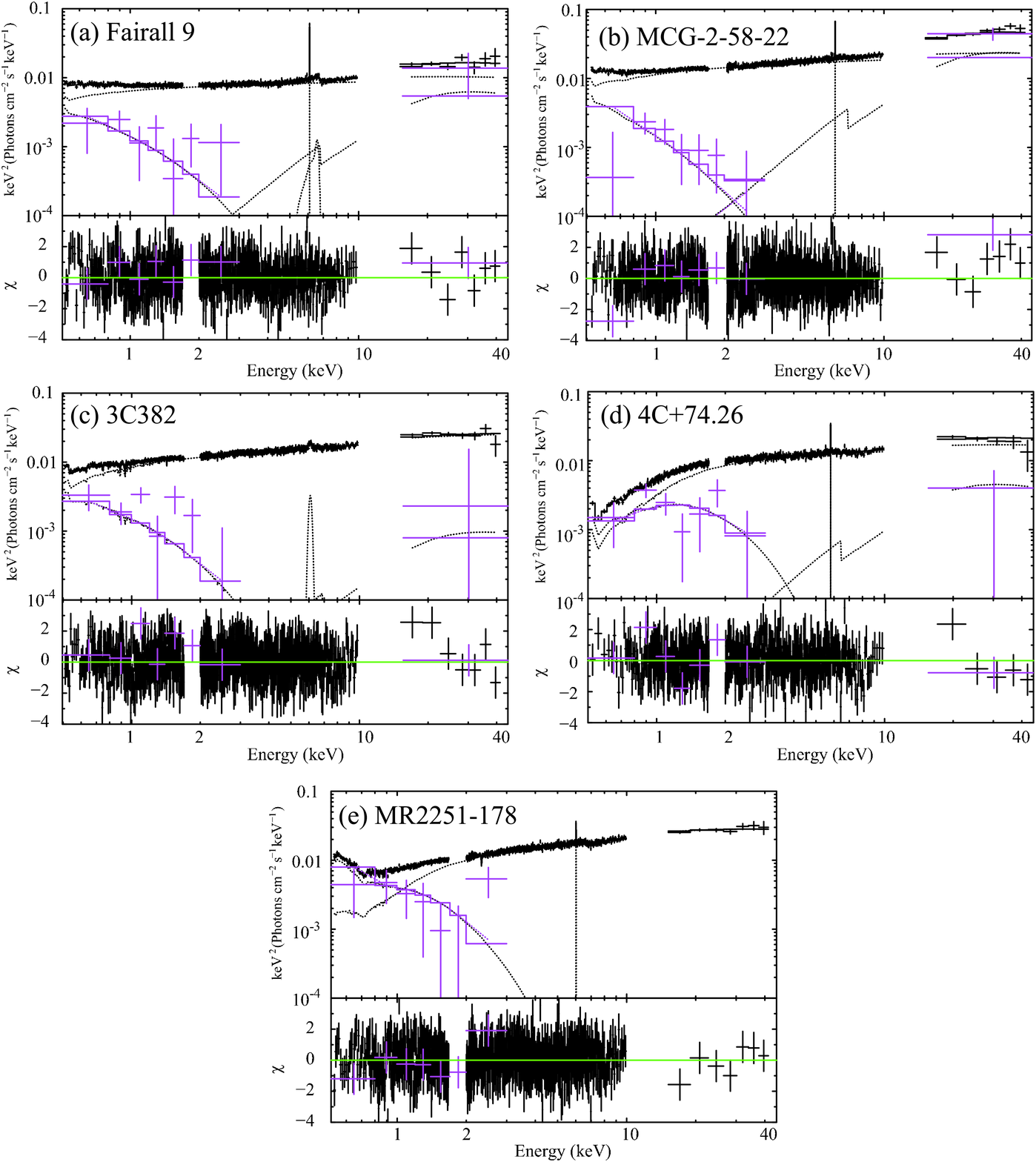}
 \end{center}
 \caption
{Same as figure 9, with the \texttt{apec} component replaced by a \texttt{comptt} model. 
The spectrum of Fairall 9 was fitted with \texttt{wabs * (cutoffpl + pexrav + zgauss + laor + comptt)},  
and that of MR2251-178 was with 
\texttt{wabs * zxipcf * (cutoffpl + pexrav + zgauss + 3 $\times$ zgauss + comptt)}, 
while those of the others were with \texttt{wabs * zxipcf * (cutoffpl + pexrav + zgauss + comptt)}. }

\label{fig:lcs}
\end{figure*}


\begin{table*}[t]
 \caption{The same as table 5, but the soft excess is represented by thermal Comptonization component (\texttt{comptt})  instead of \texttt{apec}.}
 \label{all_parametar}
 \small
 \begin{center}
  \begin{tabular}{ccccccc}
   \hline\hline
   Model & Parameter & Fairall 9 & MCG-2-58-22 &  3C382 & 4C+74.26 &MR2251-175 \\

   \hline
  \texttt{wabs} & $N_{\rm H}$
   		& $0.032$(fix)
		&$0.035$(fix)
		&$0.074$(fix)
		&$0.24^{+0.02}_{-0.02}$
		&$0.028$(fix)\\

 \texttt{zxipcf} & $N_{\rm H}$
   		           & --
		            &$0.36^{+0.24}_{-0.27}$
		            &$0.41^{+0.32}_{-0.27}$
		            &$1.08^{+0.39}_{-0.35}$
		            &$0.91 \pm 0.05$\\
		            
		             &log$\xi$
		             &--
		             &$0.36^{+0.84}_{-0.25}$
		             &$3.20^{+0.16}_{-0.13}$
		             &$0.57^{+0.18}_{-0.22}$
		             &$0.30^{+0.04}_{-0.03}$ \\
		             
		             &Cvr frac.
		             &--
		             &$0.13^{+0.21}_{-0.08}$
		             &$0.71^{+0.17}_{-0.48}$
		             &$0.54^{+0.08}_{-0.06}$
		             &$0.83^{+0.03}_{-0.02}$\\[1.5ex]

  powerlaw     & $\Gamma$%
                       & $1.87^{+0.04}_{-0.03}$
                       & $1.84^{+0.04}_{-0.02}$
                       &$1.75 \pm 0.01$
                       &$1.84 \pm 0.06$
                       &$1.66 \pm 0.01$\\

                    & $N_\mathrm{PL}$%
                       & $0.68^{+0.03}_{-0.01}$
                       & $1.33 \pm 0.01$
                       &$1.03 \pm 0.01$
                       &$0.99 \pm 0.09$
                       &$1.04^{+0.01}_{-0.03}$\\[1.5ex]
                       
   \texttt{pexrav}  & $f_\mathrm{ref}$%
                          &$0.9^{+0.3}_{-0.4}$
                          &$1.6 \pm 0.2$
                          &$<0.3$
                          &$0.4^{+0.4}_{-0.2}$
                          &$<0.1$\\[1.5ex]

 Fe~I~K$\alpha$   & $E_\mathrm{c}$

                        & $6.37^{+0.05}_{-0.04}$
                        & $6.26^{+0.08}_{-0.09}$
                        &$6.43 \pm 0.03$
                        &$6.41 \pm 0.04$
                        &$6.40 \pm 0.04$\\
                                
                                   & $\sigma$~(keV)
                               &$10^{-4}$(fix)
                               &$10^{-4}$(fix)
                           	&$0.099^{+0.032}_{-0.031}$
			&$10^{-4}$(fix)
			&$10^{-4}$(fix)\\

                   & $N_\mathrm{Fe}$ %
                        & $2.30^{+0.21}_{-0.11}$
                        & $1.89 \pm 0.32$
                        &$2.18^{+0.41}_{-0.42}$
                        &$1.46^{+0.34}_{-0.35}$
                        &$1.12^{+0.29}_{-0.30}$\\                    

                    & $EW$ (eV) %
                        & $84^{+12}_{-15}$
                        & $34^{+54}_{-32}$
                        &$48^{+32}_{-44}$
                        &$34^{+20}_{-10}$
                        &$21^{+15}_{-21}$\\[1.5ex]

\texttt{laor} & $E_\mathrm{c}$
		&$6.37 \pm 0.06$
		&--
		&--
		&--
		&--\\
		
		&$R_{\rm in}$
		&$16.2^{+22.8}_{-7.9}$
		&--
		&--
		&--
		&--\\
		
		& $N_\mathrm{Fe}$
		&$2.01^{+0.47}_{-0.50}$
		&--
		&--
		&--
		&--\\

                    & $EW$ (eV) %
                        & $91^{+51}_{-65}$
                        &--
                        &--
                        &--
                        &--\\[1.5ex]

\texttt{comptt}  & $T_{0}$~(keV)
		 &\multicolumn{5}{c}{0.02~\rm{(fix)}}\\
                        
                   &$kT$~(keV)     
                        &\multicolumn{5}{c}{0.5~\rm{(fix)}}\\      
                        
                     &$\tau$   
                        &$13.7^{+0.79}_{-0.95}$
                        &$16.2^{+1.4}_{-1.5}$
                        &$15.6^{+0.44}_{-0.31}$
                        &$18.5^{+2.07}_{-1.30}$
                        &$13.7^{+0.42}_{-0.54}$\\   
                        
                      &$N_{\rm Comp}^{*}$                                                     
                        & $0.39^{+0.08}_{-0.09}$
                        & $0.66^{+0.15}_{-0.12}$
                        &$2.27^{+0.34}_{-0.57}$
                        &$3.25^{+2.35}_{-2.17}$
                        &$42.2^{+26.7}_{-12.3}$\\ \hline

   $\chi^{2}$/d.o.f.&   & 721.33/652  & 825.32/833 & 894.07/843 & 624.26/602& 929.15/866\\
      \hline\hline
      
  \end{tabular}
 \end{center}
   
     	{\small
           \footnotemark[$*$] The thermal Comptonization component normalization, in units of 
         photons~keV$^{-1}$~cm$^{-2}$~s$^{-1}$~at 1 keV.
         }

\end{table*}


In order to improve these spectral fits, we tried adding a soft excess model,   
expressed either by a thin-thermal plasma model, an ionized and relativistically-blurred reflection model,  
or a thermal Compton model, 
which can represent the SSE (subsection 4.1). 
Because hot plasma emission in the host galaxies is generally expected in the soft X-ray band of AGN signals,  
we first employed the thin-thermal plasma model, \texttt{apec}, 
and revised the fitting model to a form of 
\texttt{wabs * zxipcf * (cutoffpl + pexrav + zgauss + apec)}. 
As before, Fairall 9 and MR2251-178 ware fitted with 
\texttt{wabs * (cutoffpl + pexrav + zgauss + laor + apec)}, and 
\texttt{wabs * zxipcf * (cutoffpl + pexrav + zgauss + 3 $\times$ zgauss + apec)}, respectively. 
In these fits, the temperature, abundance and normalization of \texttt{apec} were all left free. 
Then, as shown in table 5 and figure 9, 
the time-averaged spectra of Fairall 9, MCG-2-58-22, 3C382, and 4C+74.26 were reproduced successfully, 
while that of MR2251-178 was not, mainly due to residuals in the soft X-ray band. 
Furthermore, even in the successful four objects,  
the obtained abundances (table 5)  are much lower than the solar value,
making the fits physically unrealistic.
Thus, the thin-thermal emission model, which was allowed by the SSE modeling, 
becomes unsuccessful when the time-averaged spectra are considered.

As a next attempt, we replaced the \texttt{apec} model with an ionized reflection model, 
\texttt{reflionx}, but first without relativistic blurring, and tried to reproduce the time-averaged spectra with 
\texttt{wabs * zxipcf * (cutoffpl + pexrav + zgauss + reflionx)}, 
except for Fairall 9 which was fitted with 
\texttt{wabs * (cutoffpl + pexrav + zgauss + laor + reflionx)}, 
and MR2251-178 with 
\texttt{wabs * zxipcf * (cutoffpl + pexrav + zgauss + 3 $\times$ zgauss + reflionx)}. 
The ionization parameter and normalization of \texttt{reflionx} were left free, 
while the photon index of its primary continuum was tied to that of \texttt{cutoffpl}. 
Because the ionized reflection model is inevitably accompanied by a strong hard X-ray hump even 
when $\xi$ is made very large (e.g., $>10^3$), the spectral soft excess was not fully accounted for 
by the \texttt{reflionx} model. Specifically, 
the fits were unsuccessful except for 4C+74.26,  
with $\chi^2$/d.o.f. of 1083.54/657, 1049.98/842, 1206.17/846, 
692.57/603, and 1161.51/866 for Fairall 9, MCG-2-58-22, 
3C382, 4C+74.26, and MR2251-178, respectively. 
The fits are thus even worse than those with the \texttt{apec} modeling.  
Given this, 
the \texttt{reflionx} component in the above fits were convolved 
with the relativistic smearing kernel, \texttt{kdblur}, with  the emissivity index 
and the inner disk radius left free, 
while the outer radius fixed at 400 $R_{\rm g}$. 
Then, as summarized in table 6 and shown in figure 10, 
the fits to the time-averaged spectra of all the five AGNs were much improved 
and became acceptable for four sources, except MR2251-178.  

In the above successful fits to the 4 sources (figure 10), 
the soft-band ($<2$ keV) excess is explained by the \texttt{kdblur * reflionx} component, 
while that in the hard band by \texttt{kdblur * reflionx + pexrav}. 
Furthermore, the PL continua of the 4 sources (table 6), 
against which these soft and hard excesses are defined, 
are found to be consistent with their High-Low difference spectra 
(the main variable component; table 4, figure 8). 
Therefore, the two reflection models, 
jointly describing the static excess above the PL, 
should also be able to explain their SSE and SHE data points 
which were determined by out dynamical C3PO analysis. 
To examine this idea, 
we compared the \texttt{kdblur * reflionx + pexrav} components 
in figure 9 with the SSE+SHE data points (in purple), 
in terms of chi-square evaluation but without any parameter readjustment. 
As a result, good agreements were obtained in Fairall 9 and 3C382, 
with $\chi^2$/d.o.f. of 4.95/8 and 11.14/8, respectively, 
while poor results in MCG-2-58-22 (19.79/8) and 4C+74.26 (23.50/8). 
Thus, the \texttt{kdblur * reflionx + pexrav} modeling failed to explain the time-averaged spectrum of 
one source (MR2251-178), and caused inconsistency between the static and dynamic results 
on two more sources (MCG-2-58-22 and 4C+74.26). 

Finally, we replaced the \texttt{kdblur * reflionx} model 
with the thermal Comptonization model, \texttt{comptt}, 
and fitted the time-averaged spectra with a model of 
\texttt{wabs * zxipcf * (cutoffpl + pexrav + zgauss + comptt)}, 
except for Fairall 9 which was fitted with 
\texttt{wabs * (cutoffpl + pexrav + zgauss + laor + comptt)}, 
and MR2251-178 with 
\texttt{wabs * zxipcf * (cutoffpl + pexrav + zgauss + 3 $\times$ zgauss + comptt)}. 
In these fits, 
the disk temperature was fixed at 0.02 keV as before, 
and the electron temperature of \texttt{comptt} at 0.5 keV, 
following  Noda et al. (2011b) and Mehdipour et al. (2011),  
while its optical depth and normalization were left free. 
Like in the \texttt{apec} modeling (figure 9), 
any hard-band excess above the PL is now represented solely by cold reflection (\texttt{pexrav}). 
As summarized in table 7 and shown in figure 11, 
all the fits are acceptable, and the fit goodness is significantly (except 3C382) better than those in table 6. 
Thus, 
the soft excess component, identified statically in each object using its time-averaged spectrum, 
can be reproduced by the Comptonozation model. 
Furthermore in these fits,  
the \texttt{comptt} component has a very similar shape and intensity to the SSE spectrum. 
Actually, when the best-fit \texttt{comptt} model is compared (again without any parameter readjustment) 
with the SSE data points, we obtain $\chi^2$/d.o.f. of 4.01/7 for Fairall 9, 
10.02/7 for MCG-2-58-22, 9.52/7 for 3C382, 9.51/7 for 4C+74.26, and 5.97/7 for MR2251-178, which are 
all acceptable. 
Therefore, the soft excess structure in the time-averaged spectrum and the dynamically extracted SSE spectrum 
can be regarded as two different manifestations of the same spectral component, 
which is considered to arise via thermal Comptonization process.

Similarly, the SHE flux point of four sources (except MR2251-178)
appear to be consistent with the {\tt pexrav} component.
In fact, these SHE data points for Fairall 9 and MCG-2-58-22
have +2.1 and +5.7 sigma significances above zero, respectively,
while they deviate from the {\tt pexrav} model predictions
by 1.0 and 2.8 sigmas, respectively.
Similarly, the SHE data points of 3C 382 and 4C+74.26
(which are both statistically insignificant)
deviate from their {\tt pexrav} predictions 
by only 0.1 and 0.8 sigmas, respectively.
Thus, the dynamically derived SHE can be identified
(within the large errors) with the {\tt pexrav} model
required by the time-averaged spectrum.

\section{Discussion}

To study the origin of soft excess phenomena, 
five AGNs with different types were selected from the Suzaku archive, 
and analyzed; Fairall 9 (Sy1), MCG-2-58-22 (Sy1.5), 
3C382 (BLRG), 4C+74.26 (RLQ), and  MR2251-178 (RQQ), 
which have X-ray signals dominated by emission from their central engines. 
Applying the C3PO method developed in Noda et al. (2011b) and 
Churazov, Gilfanov, and Revnivtsev (2001), who detected  
an SSE in the bright Sy1 Mrk 509 and the typical BHB Cyg X-1, respectively,  
we succeeded in extracting an SSE in the 0.5--3 keV band of all the five AGNs. 
In addition, the application of the same technique to the 15--45 keV HXD-PIN data 
allowed us to detect the SHE from at least two of the five AGNs. 

The highly linear CCPs we obtained rule out possibilities that the soft X-ray variations of our sample AGNs 
are caused by changes in any absorption, including in particular a partially covering warm absorber 
which is one of the three main interpretations explaining 
the soft excess phenomena of AGNs (section 1). 
This is because a partially covering absorber, if variable in its covering fraction, column density, or ionization degree, 
would cause complex spectral changes, 
so the CCP would not show such a linear distribution as in figure 2--6. 
Furthermore, in this case the soft X-ray count would vanish, without leaving a positive offset, 
when the PL intensity becomes zero. Thus, the SSEs must be more independent 
of the PL continuum. 

The SSE spectra, extracted from the five AGNs via the C3PO analysis, were reproduced successfully by five  
spectral models; PL, \texttt{diskbb}, \texttt{apec}, \texttt{kdblur * reflionx} and \texttt{comptt}. 
From physical considerations, two of them, the PL and \texttt{diskbb}, were ruled out. 
Combining the static analysis with the dynamical one, 
\texttt{apec} was found to be unrealistic.  
Furthermore, as given in figure 12, the measured 0.5--3 keV luminosity of the SSE of our AGNs
is too high for thermal emission from their host galaxies 
(which would be at most $\lesssim 10^{43}$ erg s$^{-1}$; e.g., Fukazawa et al. 2006), 
and correlates positively with the AGN luminosity. 
Therefore, the SSE must be tightly connected to the AGN phenomenon, rather than to the host galaxies. 

\begin{figure*}[t]
 \begin{center}
   \FigureFile(110mm,110mm)
    {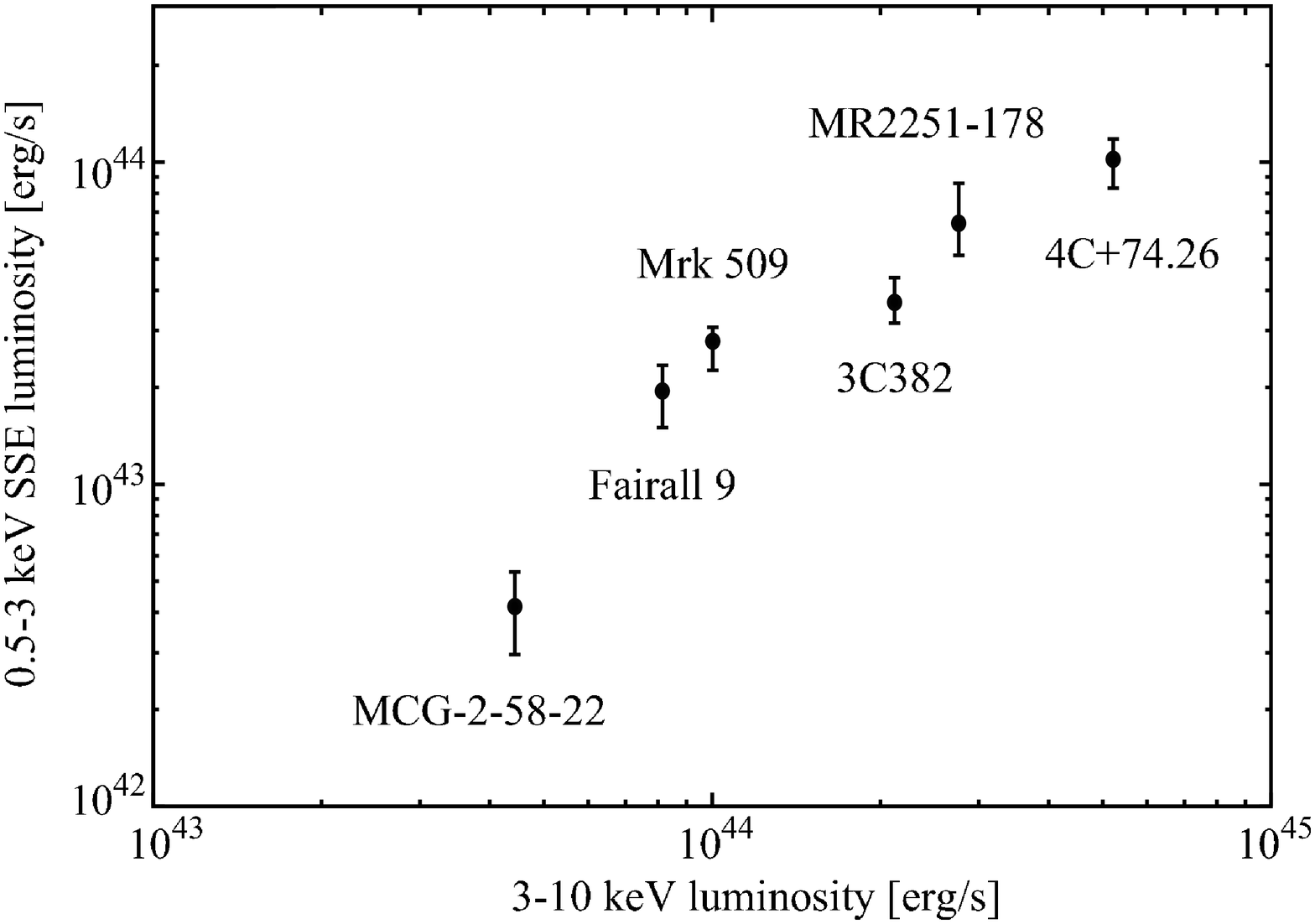}
 \end{center}
 \caption
{The 0.5--3 keV luminosity of the SSE component (table 3) of the five AGNs, 
plotted against their 3--10 keV luminosity. 
Those of Mrk 509 reported in Noda et al. (2011b) are also calculated and plotted. }

\label{fig:lcs}
\end{figure*}

How about the relativistically-smeared and ionized reflection interpretation modeled by \texttt{kdblur * reflionx}? 
When a fine-tuned geometry is actually realized to enable the ``light bending'' effect (e.g., Miniutti et al. 2007), 
the reflection would not necessarily have to follow variations of the primary emission, and may remain stable like the SSE. 
To examine whether or not this interpretation is appropriate, 
the broad energy coverage with Suzaku is particularly important, 
because the reflection signals from an ionized disk are expected 
to appear both at the softest and hardest spectral ends.
When figure 10 is closely examined,
we find that the model invoking {\tt kdblur * reflionx + pexrav} tends to
under-predict the $<3$ keV portion of the time-averaged data, 
and over-predict the time-averaged HXD data.
This effect indeed made the fit to MR2251-178 unacceptable.
Furthermore, at least in MCG-2-58-22 and 4C+74.26,
the {\tt kdblur * reflionx + pexrav} component determined through our static analysis
was not able to explain the dynamically determined SSE+SHE signals.
Thus, the relativistic reflection interpretation cannot 
explain the broad-band Suzaku data of three out of the 5 AGNs.
In contrast, the thermal Comptonization interpretation, using {\tt comptt}, 
can consistently explain, in all the 5 AGNs,
both the dynamically derived SSE and the static soft excess.
This agreement between the two independent methods significantly 
strengthens the determination of the soft excess signals (or the SSE) in our sample AGNs. 
In this case, the SHE can be explained as a separate component
arising from a cold reflector that is most likely located 
at a considerable distance from the central engine.

Presuming that the SSE is produced as a thermal Comptonization component 
which is separate from (but related to) the dominant PL, 
the corona of each AGN is then considered to consist of multiple regions 
having different optical depths and/or temperatures,
namely, different Compton y-parameters. 
This may be called Multi-Zone Comptonization (MZC) condition. 
Since the various types of AGNs which have been selected in the
present paper all have X-ray signals dominated by non-jet continua,
their central engines (where most of the gravitational energy is 
converted to radiation) are inferred to be in the MZC condition.
This agrees with the previous results on Mrk 509  (Noda et al. 2011b; Mehdipour et al. 2011), 
and the leading BHB Cyg X-1 (e.g., Makishima et al. 2008, Yamada  2011).
The lack of significant short-term ($<$ several hundred ks) variability in the SSE 
may be explained if the SSE-producing Compton corona is largely ($>$ several hundred gravitational radii) extended, 
or more likely, if the seed-photon flux (presumably from the disk) is stable on these time scales 
while the dominant PL variations are produced in those part of the corona 
with the largest y-parameter (Makishima et al. 2008). 
Overall, we suggest that the soft excess emission of 
some (if not all) AGNs is actually a part of its primary continuum,
produced in grossly the same central engine as the PL component,
but possibly at somewhat different locations 
considering the clear difference in their variation characteristics.
In this interpretation, the SHE is considered to have a different origin from the SSE, 
and produced via reflection by distant cool materials as represented by \texttt{pexrav}. 

The present results have impacts not only on the central engine,
but also on the determination and interpretation 
of various secondary spectral components,
including disk reflection, iron lines, and warm absorbers.
This is because the MZC condition affects 
the primary continuum shape,
which was often  assumed conventionally as a single PL. 
For example, Noda et al. (2011a) found 
that  the iron K$\alpha$-line width of the type I Seyfert,  MCG--6-30-15, 
decreases considerably,
when including a hard  component,
which varied independently of the dominant PL 
and made the primary continuum concave.

We thank all members of the Suzaku hardware and software teams and the Science Working Group. 
HN, HU, and SS are supported by Japan Society for the Promotion of Science (JSPS) 
Research Fellowship for Young Scientists.
KM is supported by Grantin-Aid for Scientific Research (A) (23244024) from JSPS, 
and SY by the Special Postdoctoral Researchers Program in RIKEN.

\end{document}